\documentclass[preprint, superscriptaddress,amsmath, nofootinbib]{revtex4}
\usepackage{graphicx}
\usepackage{latexsym}
\usepackage{slashed}
\usepackage{bm}
\usepackage{color}
\usepackage[colorlinks,citecolor=blue,urlcolor=blue,linkcolor=blue]{hyperref}
\usepackage{diagbox}

\def\({\left(}
\def\){\right)}

\def\beq{\begin{equation}}
\def\eeq{\end{equation}}

\begin{document}

\title{Detection of Inelastic Dark Matter via Electron Recoils in SENSEI}
	
\author{Yuchao Gu}
\email{guyc@njnu.edu.cn}
\affiliation{Department of Physics and Institute of Theoretical Physics, Nanjing Normal University, Nanjing, 210023, China}

\author{Lei Wu}
\email{leiwu@njnu.edu.cn}
\affiliation{Department of Physics and Institute of Theoretical Physics, Nanjing Normal University, Nanjing, 210023, China}

\author{Bin Zhu}
\email{zhubin@mail.nankai.edu.cn}
\affiliation{Department of Physics, Yantai University, Yantai 264005, China}

\begin{abstract}
The low-threshold experiment SENSEI, which uses the ultralow-noise silicon Skipper-CCD to explore light dark matter from the halo, has achieved the most stringent limitations on light DM-electron scattering cross section. In this work, we investigate the inelastic dark matter (IDM)-electron scattering process via the SENSEI data and derive the constraints on the inelastic dark matter model with a $U(1)$ gauge boson as the mediator. Comparing with elastic scattering process, we find that the down-scattering process with the mass splitting  $\delta \equiv m_{\chi_2}-m_{\chi_1}<0$ is more strongly constrained while the up-scattering process $\delta>0$ gets the weaker limits. For the down-scattering process with mass splitting $\delta \sim -5$ eV, the DM mass $m_{\chi}$ can be excluded down to as low as 0.1 MeV.
\end{abstract}

\maketitle

\tableofcontents

\newpage
\section{Introduction}

Although many cosmological and astrophysical observations provide strong evidence for the existence of dark matter(DM)~\cite{Trimble:1987ee}, the nature of dark matter is still elusive. Weakly Interacting Massive Particle (WIMP) with naturally correct thermal relic density is considered as a promising dark matter candidate. And yet, numerous dedicated experiments, which are aimed at directly searching for WIMP dark matter through nuclear recoil signals, have no unambiguous discoveries for WIMP~\cite{Schumann:2019eaa}. This motivates a significant reconsideration of dark matter candidates such as Light Dark Matter(LDM) with mass from keV to GeV. In recent decades, the searches for light dark matter have received a great amount of attention.

Remarkably, the nuclear recoil energy $E_{R}$ generated by light dark matter elastically scattering off target materials is typically at $E_{R} \sim \mathcal{O}(\rm{eV})$. These nuclear recoil signals are too below the thresholds of traditional direct detection experiments to be detected. Fortunately, many low threshold processes including the bremsstrahlung process~\cite{Kouvaris:2016afs,GrillidiCortona:2020owp}, Migdal effect~\cite{Ibe:2017yqa,Dolan:2017xbu,Bell:2019egg,Essig:2019xkx,GrillidiCortona:2020owp,Knapen:2020aky,Flambaum:2020xxo,Bell:2021zkr,Wang:2021oha} and electron ionization/excitation~\cite{Essig:2011nj,Bloch:2020uzh,Gao:2020wer,Ge:2020yuf,Athron:2020maw,Su:2020zny,Cao:2020bwd,An:2020bxd,Zu:2020bsx,Guo:2020oum,Du:2020ybt,Chao:2021liw,Chen:2021ifo} will occur with scattering process and lead to visible signals. Moreover, thanks to the developments of direct detection technologies such as single-electron detection technology, these tiny photo-electric signals can be probed by detectors, allowing for the detection of LDM. Apart from the advanced detection technology, low threshold materials are also extensively investigated to search for LDM, including superconductors~\cite{Hochberg:2015pha,Hochberg:2015fth,Hochberg:2019cyy}, silicon and germanium semiconductors~\cite{Essig:2011nj,Graham:2012su,Lee:2015qva,Essig:2015cda,Crisler:2018gci,SuperCDMS:2018mne,SENSEI:2019ibb,CDEX:2019hzn,DAMIC:2019dcn,Andersson:2020uwc,SENSEI:2020dpa,SuperCDMS:2020ymb,Catena:2021qsr}, argon~\cite{DarkSide:2018ppu} and xenon~\cite{Essig:2011nj,Essig:2012yx,Essig:2017kqs,PandaX-II:2017hlx,XENON:2019gfn} noble liquids, scintillators~\cite{Derenzo:2016fse,Blanco:2019lrf}, graphene two-dimensional materials~\cite{Hochberg:2016ntt,Geilhufe:2018gry}, three-dimensional Dirac materials~\cite{Hochberg:2017wce,Geilhufe:2019ndy}, polar crystal~\cite{Knapen:2017ekk}.

On the other hand, Light dark matter that naturally satisfies the current observed DM relic density can appear in many prevalent dark matter models~\cite{Boehm:2003ha,Borodatchenkova:2005ct,Pospelov:2007mp,Hooper:2008im,Feng:2008ya,Pospelov:2008jk,Falkowski:2011xh}. An interesting model is the inelastic dark matter(IDM) model~\cite{Tucker-Smith:2001myb,Tucker-Smith:2004mxa,Finkbeiner:2007kk,Arina:2007tm,Chang:2008gd,Cui:2009xq,Lin:2010sb,DeSimone:2010tf,An:2011uq,Pospelov:2013nea,Finkbeiner:2014sja,Dienes:2014via,Dror:2019onn,Dror:2019dib,CarrilloGonzalez:2021lxm,Guo:2021vpb}, initially motivated by reconciling the tension between DAMA and CDMS data. Usually, the dark sector in the inelastic dark matter model interacts with SM particles via a new gauge boson charged under the extra $U(1)_{D}$ gauge symmetry. Different from DM-electron elastic scattering processes, the IDM-electron scattering will bring richer phenomenology. Furthermore, the recent XENON1T electron excess can be also addressed by inelastic dark matter~\cite{Harigaya:2020ckz,Lee:2020wmh,Baryakhtar:2020rwy,Bramante:2020zos,Choi:2020ysq,Emken:2021vmf,Dror:2020czw} with the mass splitting $\delta \sim 2-3$ keV at which the peak of the electron recoil energy spectrum excess occurs. Besides, the inelastic dark matter models have been widely studied in DM direct detection experiments~\cite{An:2020tcg,Song:2021yar,Bell:2021zkr,Bell:2021xff}. 

In this paper, we will investigate inelastic dark matter scattering off electrons bound to semiconductors by exploiting the latest released data from the Sub-Electron-Noise Skipper-CCD Experimental Instrument (SENSEI)~\cite{SENSEI:2020dpa}. The SENSEI experiment, located at deep underground in the MINOS cavern at Fermi National Accelerator Laboratory(FNAL), is able to probe DM mass down to $m_{\chi} \sim$ 0.5 MeV by using the ultralow-noise silicon Skipper-Charge-Coupled-Devices. Given the lower binding energy of semiconductors $E_{\rm{gap}} \sim \mathcal{O}$(eV), the lower DM mass could be explored by the combination of low threshold target materials and inelastic dark matter features. The IDM-electron scattering will excite electrons from a valence band to a conduction band, resulting in the observable electron-hole pairs $N_{e}$. 
We derive the 90$\%$ confidence level (C.L.) exclusion limits on DM-electron cross section $\bar{\sigma}_{e}$ by using the observed number of electron-hole pairs $N_{e}$ from the SENSEI experimental data. 

The paper is organized as follows. In Sec.~\ref{sec2}, we introduce the inelastic dark matter model with a new $U(1)_{D}$ gauge boson, mediating the DM particles and electrons interaction. Besides, we also evaluate the events induced by IDM-electron scattering, including the calculations of kinematical and dynamical processes as well as the crystal form factor $f_{c}(q,\Delta E_{e})$. In Sec.~\ref{sec3} we exhibit the generated events as a function of the electron deposited energy $\Delta E_{e}$ with different DM masses $m_{\chi}$, mass splitting $\delta$ and DM form factors $F_{\rm DM}$. Also, we obtain the 90$\%$ C.L. bounds on $\bar{\sigma}_{e}-m_{\chi}$ plane by utilizing the SENSEI exprimental data. Finally, we draw some conclusions in Sec~\ref{sec4}.

\section{Inelastic Dark Matter-Electron Scattering in Semiconductor}
\label{sec2}

We introduce a Dirac fermion dark matter $\chi$ in the model where the dark sector interacts with SM particles through a new mediator dark photon $A_{\mu}^{'}$, which is kinetically mixing with the photon~\cite{Holdom:1985ag}
\begin{equation}
    \mathcal{L} \supset \frac{1}{2} \epsilon F^{\mu \nu} F_{\mu \nu}^{'},
    \label{kinetic mixing}
\end{equation}
where $\epsilon$ is the kinetic mixing parameter, $F^{\mu \nu} \left(F_{\mu \nu}^{'}\right)$ is the QED strength field tensor (the dark photon strength field tensor). The interactions between the dark sector are governed by the lagrangian 
\begin{equation}
    \mathcal{L}_{D} \supset \bar{\chi}\left(i\slashed{D}-m_{\chi}\right)\chi-\frac{\delta}{4} \left(\bar{\chi^{c}} \chi+\rm{h.c.}\right),
\end{equation}
where the covariant derivative is $D_{\mu} \equiv \partial_{\mu} + i g_{D} A_{\mu}^{'}$ and the $g_{D}$ is the U(1)$_{D}$ gauge coupling. The dark sector communicates with the standard model electromagnetic current $\mathcal{J}^{\mu}$ by exchanging the $U(1)_{D}$ gauge boson dark photon $A^{'}_{\mu}$. Notably, the mass splitting arises from the Majorana mass term, which may be generated from the Higgs mechanism through $\chi$ and ${\chi^{c}}$ Yukawa couplings. Therefore, these dark sector interactions can decompose the Dirac fermion $\chi$ into two almost degenerate Majorana mass eigenstates
\begin{align}
    \chi_{1}=\frac{1}{\sqrt{2}}\left(\chi-\chi^{c}\right), m_{\chi_{1}}= m_{\chi}-\frac{\delta}{2}, &\\
    \chi_{2}=\frac{1}{\sqrt{2}}\left(\chi+\chi^{c}\right), m_{\chi_{2}}= m_{\chi}+\frac{\delta}{2},
\end{align}
where their mass splitting $\delta=m_{\chi_2}-m_{\chi_1}$ is much smaller than $m_{\chi}$. $\chi_{1}$($\chi_{2}$) are the ground (excited) state. Besides, the dark photon $A_{\mu}^{'}$ can receive the mass $m_{A^{'}}$ through the Stueckelberg mechanism or dark Higgs.

The dark matter particles $\chi$ inelastically scatter off electrons bound to semiconductors through exchanging the dark photon $A_{\mu}^{'}$. The transfer momentum $q$ is negligible relative to the electron mass $m_{e}$ and DM mass $m_{\chi} \sim \mathcal{O}(\rm{MeV})$ since it is typically at keV scale.
Furthermore, the IDM-electron scattering cross section is approximately equal to that of elastic scattering since the mass splitting $\delta \ll m_{\chi}$ as mentioned before. Therefore, the inelastic cross section is decomposed into the reference cross section $\bar{\sigma}_{e}$ and DM form factor $F_{\rm DM}$,
\begin{equation}
    \sigma_{\chi e}= \bar{\sigma}_{e} \left|F_{\rm DM}(q)\right|^2 = \frac{16 \pi \mu \alpha \alpha_{D} \epsilon^2}{(m_{A^{'}}^2+\alpha^2 m_{e}^2)^2} |F_{\rm DM}(q)|^2
\end{equation}
with 
\begin{align}
     F_{\rm DM}(q)= \frac{m_{A^{'}}^2+ \alpha^2 m_{e}^2}{m_{A^{'}}^2+q^2} \simeq \begin{cases}1, & m_{A^{'}} \gg \alpha m_{e}\\ \left(\frac{\alpha m_{e}}{q}\right)^2, & m_{A^{'}} \ll \alpha m_{e}\end{cases},
\end{align}
where the reference cross section $\bar{\sigma}_{e}$ is set up by the transfer momentum $q=\alpha m_{e}$ and $\mu$ is the DM-electron reduced mass. $\alpha_{D} \equiv g_{D}^2/4 \pi$ and $\alpha$ is the fine structure constant respectively. The transfer momentum $q$-dependent terms are absorbed in the DM form factor $F_{\rm DM}$. We study two scenarios where the heavy mediator $m_{A}^{'} \gg \alpha m_{e}$ and the light mediator $m_{A}^{'} \ll \alpha m_{e}$. The DM form factor $F_{\rm{DM}}=1$ is for the heavy mediator whereas $F_{\rm{DM}}=\left(\frac{\alpha m_{e}}{q}\right)^2$ is for the light mediator.

The kinematics of IDM-electron scattering, including the up-scattering ($\chi_{1} e \rightarrow \chi_{2} e$) and down-scattering processes ($\chi_{2} e \rightarrow \chi_{1} e$), satisfy the energy conservation
\begin{equation}
    \frac{1}{2} m_{\chi} v^2=\frac{|m_{\chi} \vec{v}-\vec{q}|^2}{2 m_{\chi}} + \delta + \Delta E_{e},
    \label{energy conservation}
\end{equation}
where the electron deposited energy $\Delta E_{e}$ is the energy difference between the initial and final electron energy, $\vec{q}$ is the transfer momentum and $v$ is the velocity of the incoming dark matter. Besides, $\delta$ is positive for the up-scattering process while $\delta$ is negative for down-scattering process. After simplification, the energy conservation Eq.~\ref{energy conservation} can be expressed by the following form
\begin{equation}
    \frac{q^2}{2 m_{\chi}}-\vec{q} \cdot \vec{v} + \delta + \Delta E_{e}=0.
\end{equation}
Thus, the maximum(minimum) transfer momentum $q_{\rm{max}}(q_{\rm{min}})$ can be written as
\begin{equation}
    q_{\rm{max}}=m_{\chi} v \left(1 + \sqrt{1-\frac{2(\delta+\Delta E_{e})}{m_{\chi} v^2}}\right),
\end{equation}
\begin{align}
    q_{\rm{min}}=\begin{cases}m_{\chi}v\left(1-\sqrt{1-\frac{2(\delta+\Delta E_{e})}{m_{\chi} v^2}}\right) & \delta+\Delta E_{e}>0 \\
    m_{\chi}v\left(-1+\sqrt{1-\frac{2(\delta+\Delta E_{e})}{m_{\chi} v^2}}\right) & \delta + \Delta E_{e}<0
    \end{cases},
\end{align}
with cos$\theta=1$, where $\theta$ is the angle between the incoming DM velocity $\vec{v}$ and the transfer momentum $\vec{q}$. When the limit $\delta \rightarrow 0$, the maximum(minimum) transfer momentum returns to the elastic case. For a given electron deposited energy $\Delta E_{e}$ and the transfer momentum $q$, the kinematically allowed minimum velocity of the incoming dark matter $v_{\rm{min}}$ is given by
\begin{equation}
    v_{\rm{min}}=\left|\frac{\Delta E_{e}+\delta}{q}+\frac{q}{2 m_{\chi}}\right|.
\end{equation}
In order to prevent DM from escaping the galaxy, this puts the upper limit on the minimum velocity of the incoming dark matter $v_{\rm{min}} \leq v_{\rm{E}}+v_{\rm{esc}}$, where $v_{\rm{E}}=240$ km/s is the average Earth velocity relative to the DM halo and $v_{\rm{esc}}=600$ km/s is the escape velocity of the galaxy. This constraint $v_{\rm{min}} \leq v_{\rm{E}}+v_{\rm{esc}}$ is valid for up-scattering and down scattering processes. Additionally, in the consideration of the up-scattering process, the kinetic energy of dark matter partices $E_{k}^{\chi} \sim \frac{1}{2} m_{\chi} v^2$ should be larger than the mass splitting $\delta$, which guarantees that the up-scattering process is kinematically allowed. This requirement also provides the upper limit on the mass splitting $\delta$ with the maximum
\begin{equation}
    \delta_{\rm{max}}=\frac{1}{2} q_{\rm{max}}(v_{\rm{esc}}+v_{\rm{E}})-\Delta E_{e} \sim 100 \ \rm{eV},
\end{equation}
above which the kinetic energy $E_{k}^{\chi}$ cannot compensate the mass splitting $\delta$ for the up-scattering process. It should be noted that we derive the maximum of mass splitting $\delta_{\rm max}$ with $q_{\max}=18 \alpha m_{e}$ and $\Delta E_{e}=0.1$ eV, which are mentioned in the following numerical calculation of the crystal form factor. When evaluating the events induced by IDM-electron scattering, we consider the Standard Halo Model(SHM)~\cite{Drukier:1986tm} where the local DM velocity is described by Maxwell-Boltzmann distribution. The dependence of the generated events on different galactic dark matter velocity distributions is discussed in Refs~\cite{Radick:2020qip,Maity:2020wic}. 
Assuming that the DM velocity distribution is spherically symmetric, we take the form of the Maxwell-Boltzmann distribution in the detector rest frame
\begin{equation}
    B(\vec v_\chi)=\frac{1}{N v_{0}^3 \pi}e^{-\frac{|\vec v_\chi+\vec v_E|^2}{v_0^2}} \Theta(v_{\rm{esc}}-|\vec{v}_{\chi}+\vec{v}_{E}|)
\end{equation}
with the normalization factor
\begin{equation}
    N=\sqrt{\pi} \textrm{Erf} \left(\frac{v_{\rm esc}}{v_0}\right)-2\left(\frac{v_{\rm esc}}{v_0}\right)e^{-\left(\frac{v_{\rm esc}^2}{v_0^2}\right)}
\end{equation}
where $v_{0}=230$ km/s is the typical velocity of the halo DM. The velocity dependent integral has the following expression
\begin{equation}
    \eta(q,\Delta E_{e})=\int d^3v \frac{B(\vec{v}_{\chi})}{v} \Theta(v-v_{\rm{min}}(q,\Delta E_{e})),
\end{equation}
which is eventually expressed by a piecewise function as shown in Refs~\cite{Essig:2015cda}.
The differential event rate $dR_{\rm{c}}/dE_{e}$ produced by the IDM-electron scattering in a semiconductor target is determined by
\begin{equation}
    \frac{dR_{\rm{c}}}{dE_{e}}=\frac{\rho_{\chi}}{m_{\chi}} \frac{1}{m_{\rm{T}}} \bar{\sigma}_{e} \alpha \frac{m_{e}^2}{\mu^2} \int dq F_{DM}(q)^2 |f_{c}(q,\Delta E_{e})|^2 \eta(q,\Delta E_{e}),
    \label{eqevents}
\end{equation}
where the $\rho_{\chi}=0.3$ GeV/cm$^3$~\cite{Bovy:2012tw} is the local dark matter density, $m_{T}$ is the mass of target material, and $f_{c}(q,\Delta E_{e})$ is the crystal form factor for exciting an electron from a valence band to a conduction band in semiconductors. Because there exists two silicon atoms in each silicon crystal, the target mass $m_{T}=2 m_{\rm{Si}}=52.33$ GeV. In the following, we will pay more attention to calculating the crystal form factor $f_{c}(q,\Delta E_{e})$, which is related to the overlap integral of the initial and final electron wave functions. We exploit the QEdark code~\cite{Essig:2015cda} based on Quantum ESPRESSO to numerically evaluate the crystal form factor.
Because of the periodic potential in a semiconductor crystal, electrons bound to a semiconductor valence band are governed by Bloch wave functions, $\psi_{\bf{k}}^n (\bf{r})$ with the band label $n$ and the electron momentum $\bf{k}$ in the first Brillouin Zone(BZ) 
\begin{equation}
    \psi_{\bf{k}}^n ({\bf{r}})= \frac{1}{\sqrt{V}} \sum_{\bf{G}} \psi^n({\bf{k}+\bf{G}}) e^{i({\bf{k}+\bf{G}}) {\bf{r}}}
\end{equation}
with the normalization condition
\begin{equation}
    \sum_{\bf{G}} |\psi^n ({\bf{k}+\bf{G}})|^2=1,
\end{equation}
where $V$ is the volume of the crystal and ${\bf{G}}$ is the reciprocal lattice vector. The form factor $f_{n{\bf{k}}\rightarrow n^{'}{\bf{k}^{'}},{\bf{G}^{'}}}$ related to electron excitation from a valence band ${\{n,\bf{k}}\}$ to a conduction band ${\{n^{'},\bf{k}^{'}}\}$ is described by
\begin{equation}
    f_{n{\bf{k}}\rightarrow n^{'}{\bf{k}^{'},\bf{G}^{'}}}=\sum_{\bf{G}}\psi^{n^{'}*} ({\bf{k}^{'}+\bf{G}^{'}+\bf{G}}) \psi^n(\bf{k}+\bf{G}),
\end{equation}
The crystal form factor as a function of transfer momentum $q$ and the electron deposited energy $\Delta E_{e}$ has the following expression
\begin{eqnarray}
|f_{c}(q,\Delta E_{e})|^2 &=& \frac{2 \pi^2 \left(\alpha m_{e}^2 V_{\rm{cell}}\right)^{-1}}{\Delta E_{e}} \sum_{n n^{'}} \int_{\rm{BZ}} \frac{V_{\rm{cell}} d^3k}{(2\pi)^3} \frac{V_{\rm{cell}} d^3k^{'}}{(2\pi)^3} \nonumber \\ 
&\times& \Delta E_{e} \delta(\Delta E_{e}-E_{n^{'},\bf{k}^{'}}+E_{n,\bf{k}}) \sum_{\bf{G}^{'}}{q \delta(q-|\bf{k}^{'}-\bf{k}+\bf{G}^{'}|)} \left| f_{n{\bf{k}}\rightarrow n^{'}{\bf{k}^{'},\bf{G}^{'}}}\right|^2.
\end{eqnarray}
where the crystal form factor sums over both all filled energy bands $\{n,\bf{k}\}$ and unfilled energy bands $\{n^{'},\bf{k}^{'}\}$, the transfer momentum $q$ integrates over the first BZ. The factor $2 \pi^2 (\alpha m_{e}^2 V_{\rm{cell}})^{-1}$ with the dimension of energy equals to 2.0 eV for silicon semiconductors and $V_{\rm{cell}}$ is the volume of the unit cell. Here $E_{n,\bf{k}}$($E_{n^{'},\bf{k}^{'}}$) is the energy of level $\{n,\bf{k}\}$($\{n^{'},\bf{k}^{'}\}$). These two $\delta$-functions are required by the energy and momentum conservation.

For numerically calculating the crystal form factor $f_{c}(q,\Delta E_{e})$, we use these methods described in Refs~\cite{Essig:2015cda}: binning in $q$ and $\Delta E_{e}$, discretization in $\bf{k}$ and cutoff in $\bf{G},\bf{G^{'}}$ shown in Appendix~\ref{sec6}. These operations are encoded in the QEdark code. The modifications that we made to the QEdark code are the calculations of kinematic part where the mass splitting $\delta$ is encoded in those relevant functions as described in Sec.~\ref{sec2}. In Eq.~\ref{eqevents}, the differential event rate $dR_{c}/dE_{e}$ is a function of the electron deposited energy $\Delta E_{e}$. However, the electron deposited energy cannot be directly measured by DM direct detection experiments. Instead, we should convert $\Delta E_{e}$ to $N_{e}$ since the electron-hole pairs $N_{e}$ are detectable. Because of the conversion of $\Delta E_{e}$ to $N_{e}$ involving a complicated chain of secondary scattering processes, there is no exact model describing these secondary scattering processes so far. We assume a linear response function, which is regarded as a reasonable assumption describing the true behavior
\begin{equation}
    N_{e}=1+{\rm Floor}\left[(\Delta E_{e}-E_{\rm{gap}})/\epsilon\right],
\end{equation}
where $E_{\rm{gap}}=1.2$ eV is the band energy and $\epsilon=3.8$ eV~\cite{Rodrigues:2020xpt} is the mean energy per electron-hole pair for silicon semiconductors. And the floor function Floor[$x$] represents the nearest integer less than or equal to $x$. The first term in the linear response function represents the primary electron-hole excited by the initial IDM-electron scattering, while the second term shows the additional electron-hole pairs induced by the residual electron deposited energy. Therefore, the observable number of electron-hole pairs is evaluated by
\begin{equation}
R_{N e^{-}}= \int \frac{dR}{dE_{e}} \delta \left(1+{\rm Floor}\left[(\Delta E_{e}-E_{\rm{gap}})/\epsilon\right]-N_{e}\right)dE_{e}.
\label{Neevents}
\end{equation}

Furthermore, before the dark matter particles arrive at the underground detectors, we should take into account the Earth shielding effect~\cite{Xia:2021vbz}, containing two cases. One scenario is that the halo dark matter is mainly made of the ground states $\chi_{1}$. Before the ground states $\chi_{1}$ reach the detector, they will be converted to the excited states $\chi_{2}$ via up-scattering off atoms in the Earth. This terrestrial up-scattering effect is discussed in Refs~\cite{Emken:2021vmf}, where the fraction of excited states created by up-scattering process only accounts for $\mathcal{O}(10^{-4})$. It is much smaller than the local dark matter density $\rho_{\chi}$. Therefore, the excited states generated by the Earth shielding effect can be negligible. Also, we can assume another case where the halo dark matter is fully composed by the excited states $\chi_2$ because of its long lifetime. They will de-excite to the ground states $\chi_{1}$ through down-scattering off atoms before their reaching the detectors, which results in the number density $\rho_{\chi}$ decreasing. Given the previous scenario, it is therefore reasonable to speculate that only a small fraction of excited states $\chi_{2}$ convert to the ground states through down-scattering process. The accurate calculation of the Earth shielding effect in this scenario will be delayed in the future work.
\section{Numerical Results and Discussions}
\label{sec3}
\begin{figure}[ht]
\centering
\includegraphics[height=7cm,width=8cm]{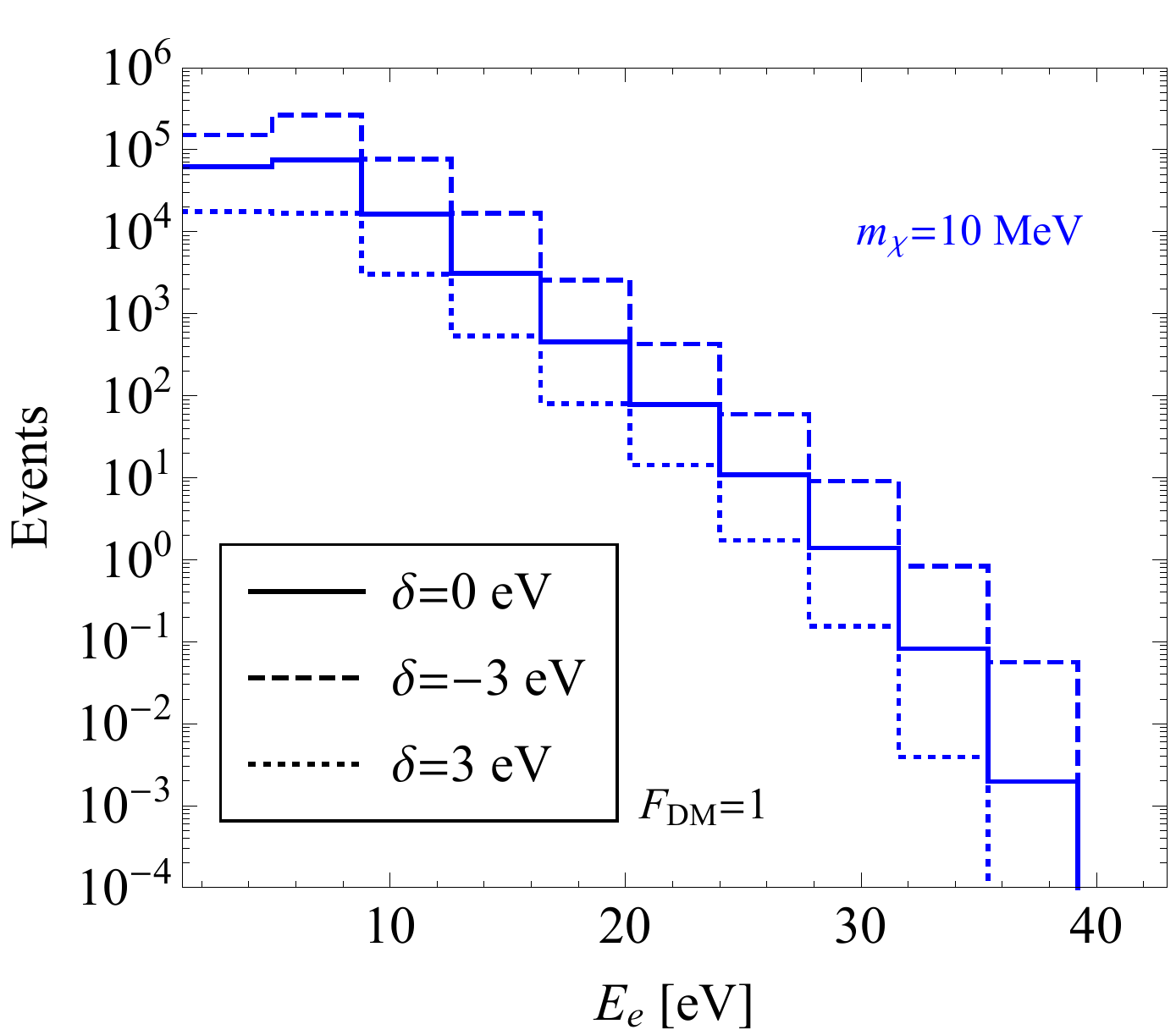}
\includegraphics[height=7cm,width=8cm]{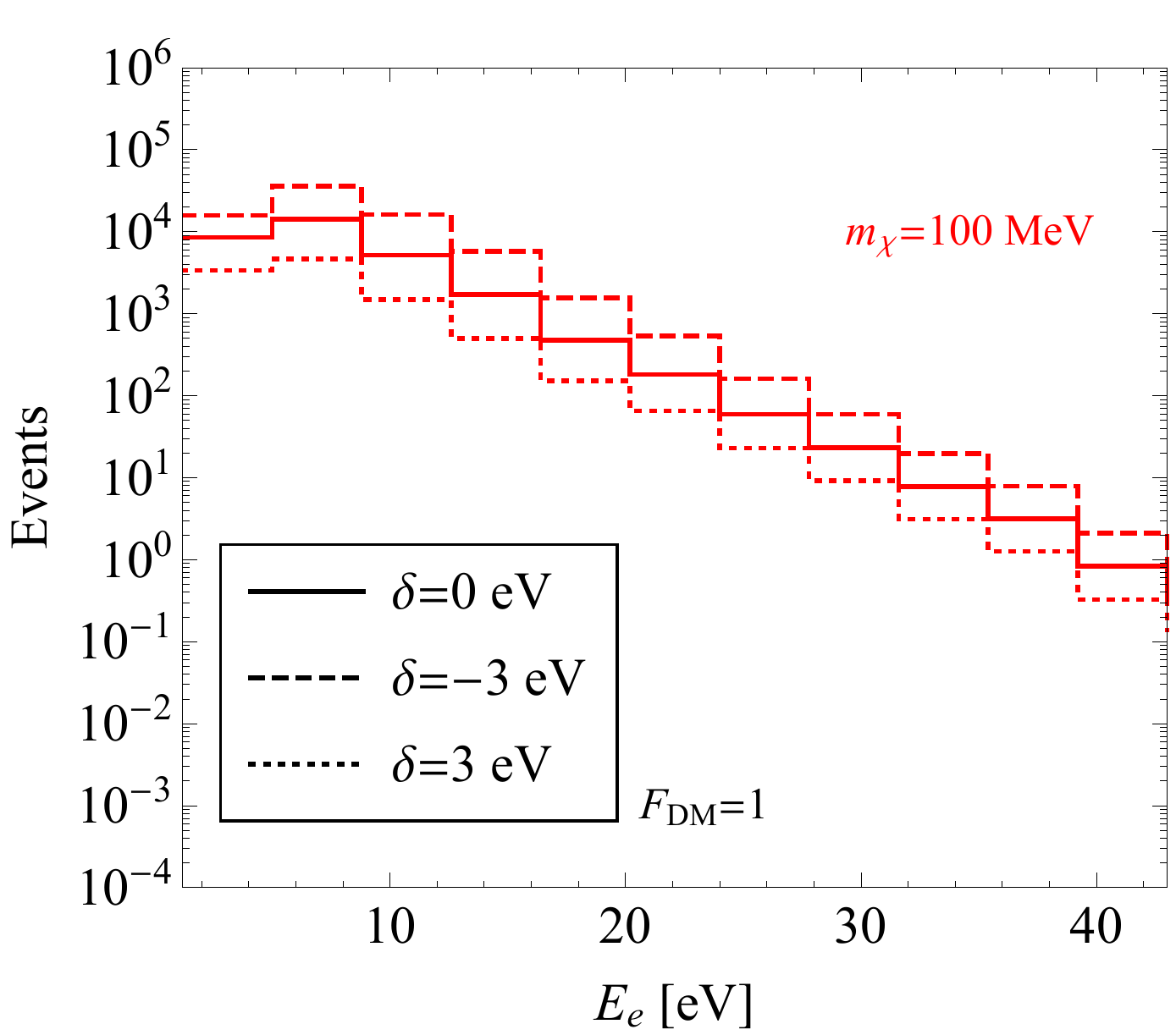}
\includegraphics[height=7cm,width=8cm]{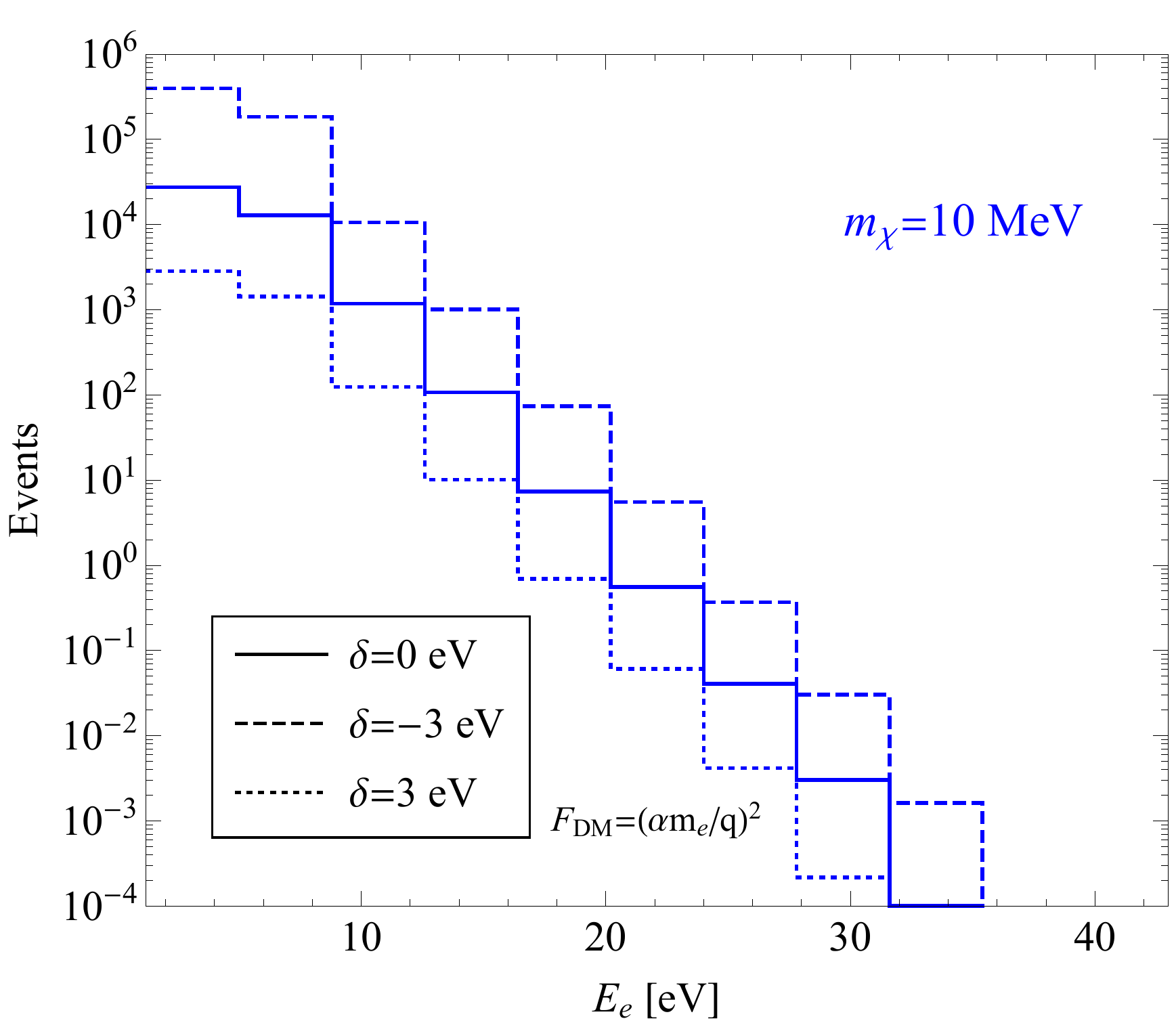}
\includegraphics[height=7cm,width=8cm]{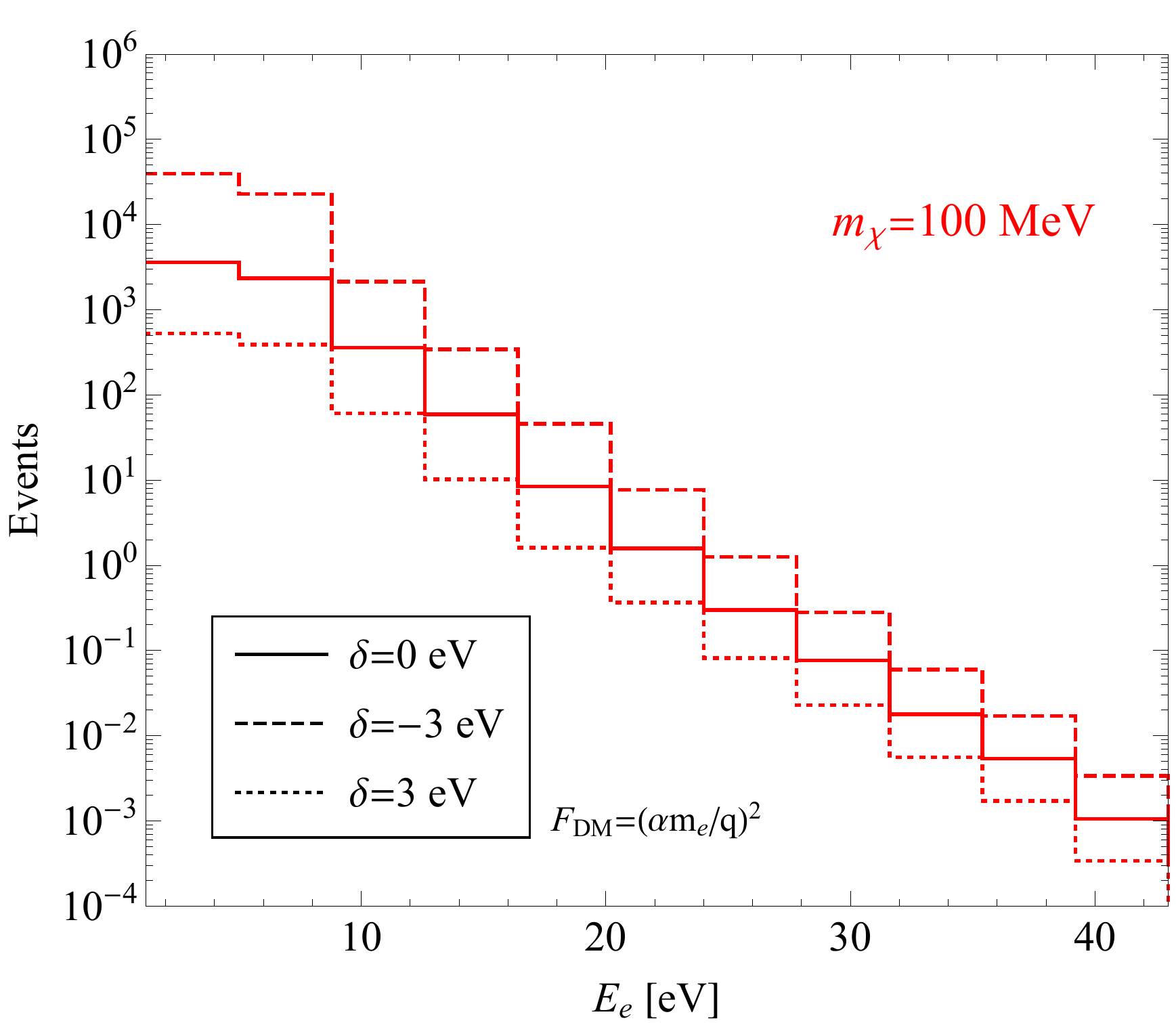}
\caption{The number of events arising from the IDM-electron scattering with the exposure 1kg$\cdot$year versus the electron deposited energy $\Delta E_{e}$ for different mass splitting $\delta$, DM masses $m_{\chi}$ and DM form factors $F_{\rm{DM}}$. The solid line represents the number of events generated by the elastic DM-electron scattering. While the dashed and dotted line show the number of events originating from IDM-electron scattering with negative($\delta=-3$ eV) and positive($\delta=3$ eV) mass splitting respectively.}
\label{events}
\end{figure}
We show the number of events induced by the IDM-electron scattering in Fig.~\ref{events} and consider two different DM form factors $F_{\rm{DM}}$. $F_{\rm{DM}}=1$ (upper panel) indicates that the dark photon $A_{\mu}^{'}$ is a heavy mediator whereas $F_{\rm{DM}}=(\alpha m_{e}/q)^2$ (bottom panel) represents that the dark photon $A_{\mu}^{'}$ is a light mediator. The events produced by two different DM masses $m_{\chi}=10$ MeV, $m_{\chi}=100$ MeV are shown with the blue and red lines. As shown in Fig.~\ref{events}, the number of events in each electron deposited energy bin decreases with the electron deposited energy $\Delta E_{e}$ increasing. This is because that the crystal form factor $f_{c}(q,\Delta E_{e})$ is highly suppressed by the large transfer momentum $q$. The large transfer momentum $q$ indicates the large $\Delta E_{e}$, resulting in the less events in large $\Delta E_{e}$ region. Note that the events mentioned here are calculated by Eq.~\ref{Neevents} rather than the experimentally observed data. Different from the elastic DM-electron scattering process, the down-scattering process $\delta<0$ produces the most events while the up-scattering process $\delta>0$ generates the least with the same DM mass $m_{\chi}$, DM form factor $F_{\rm{DM}}$ and the electron deposited energy $\Delta E_{e}$ in each picture in Fig.~\ref{events}.  The DM kinetic energy not only excites electrons but also converts to the mass splitting $\delta$ for the up-scattering process. Whereas for the down-scattering process, the mass splitting $\delta$ can also contribute the extra energy to electron excitation in addition to the DM kinetic energy. Thus, the events induced by down-scattering process are more than those produced by the up-scattering process. Beside, the light dark matter results in more events in the small $\Delta E_{e}$ region while the heavy dark matter generates more events in the large $\Delta E_{e}$ region for the same $\delta$ and $F_{\rm{DM}}$. On one hand, both $m_{\chi}=10$ MeV and $m_{\chi}=100$ MeV have enough kinetic energy to excite electrons in the small $\Delta E_{e}$ region. However, the generated events are enhanced by the DM mass $1/m_{\chi}$ as described in Eq.~\ref{eqevents}, giving rise to the more events for $m_{\chi}=10$ MeV in small $\Delta E_{e}$ range. On the other hand, although the resulting events are enhanced by $1/m_{\chi}$, the light dark matter lacks enough kinetic energy to excite more electrons in the large $\Delta E_{e}$ region. Conversely, the heavy dark matter have enough kinetic energy to induce more electron excitation in the large $\Delta E_{e}$ region, which results in more events for $m_{\chi}=100$ MeV. Additionally, the dependence of generated events on two different DM form factors $F_{\rm {DM}}$ will be displayed with same $m_{\chi}$ and $\delta$. It should be noted that the larger electron deposited energy $\Delta E_{e}$ implies the larger transfer momentum $q$ as mentioned before.  Compared with the DM form factor $F_{\rm{DM}}=1$, the induced events for $F_{\rm{DM}}=(\alpha m_{e}/q)^2$ in small transfer momentum region $\left(q< \alpha m_{e}\right)$, namely small $\Delta E_{e}$, are large due to the produced events being enhanced by $F_{\rm{DM}}$ while those in large transfer momentum region $\left(q> \alpha m_{e}\right)$ are relatively small because of the resulting events being highly suppressed by $F_{\rm{DM}} \sim 1/q^{2}$.
\\
\begin{figure}[ht]
\centering
\includegraphics[height=8cm,width=8cm]{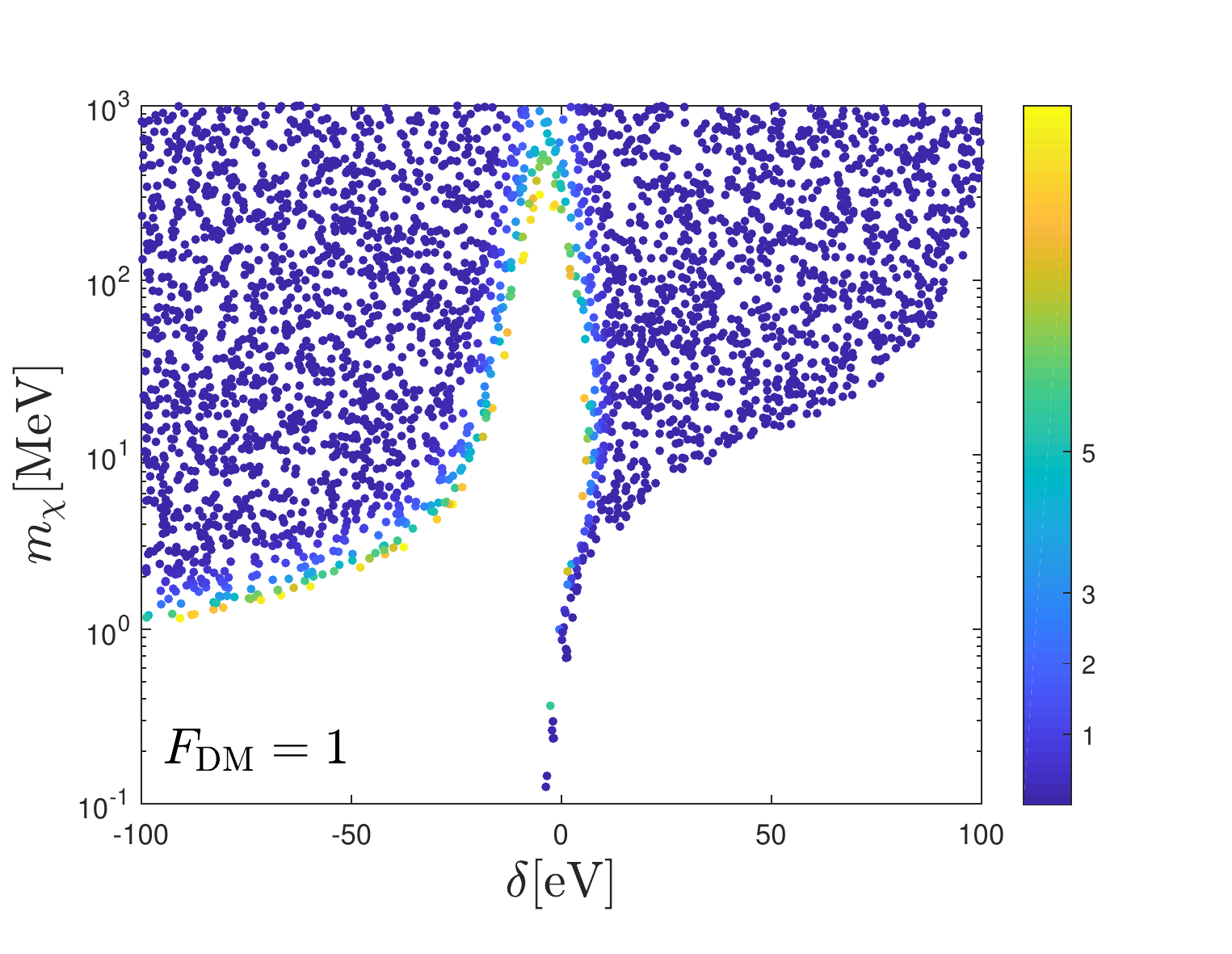}
\includegraphics[height=8cm,width=8cm]{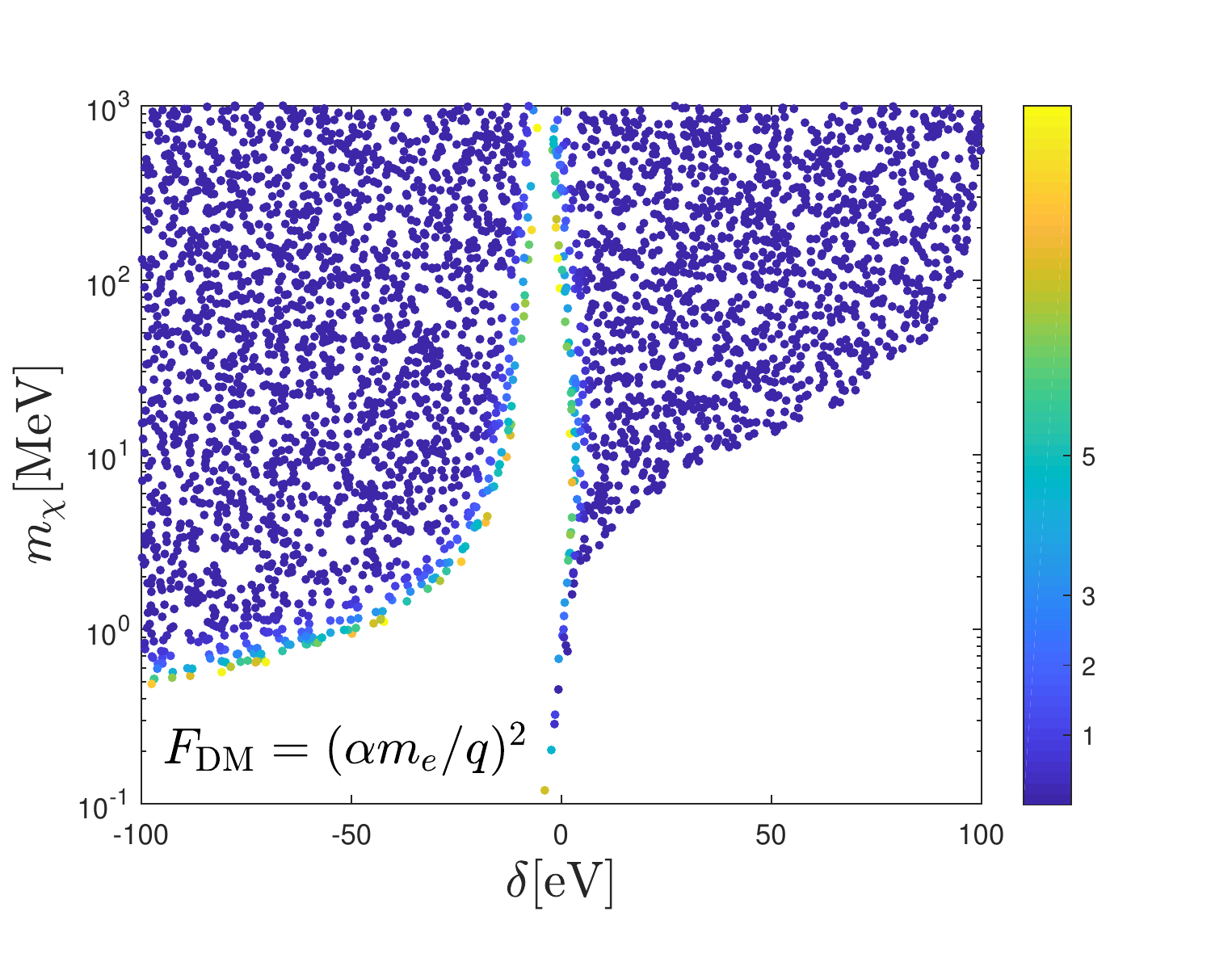}
\caption{The numerical results on $m_{\chi}-\delta$ plane are derived by using the experimentally observed number of electrons $R_{1 e^{-}}^{\rm{obs}}$ from SENSEI data with reference cross section $\bar{\sigma}_{e}=10^{-37}$ cm$^{-2}$ and two different DM form factors $F_{\rm{DM}}=1$ (left picture), $F_{\rm{DM}}=(\alpha m_{e}/q)^2$ (right picture). The color bar represents the ratio of $R_{1 e^{-}}^{\rm theory}$ to $R_{1 e^{-}}^{\rm{obs}}$, whose range is $0<r=R_{1 e^{-}}^{\rm theory}/R_{1 e^{-}}^{\rm{obs}}\leq 10$.}
\label{mdelta}
\end{figure}

In Fig.~\ref{mdelta}, we present the ratio of theoretically evaluated events $R_{1 e^{-}}^{\rm theory}$ to the experimentally observed events $R_{1 e^{-}}^{\rm{obs}}$ in the $m_{\chi}-\delta$ plane. The $R_{1 e^{-}}^{\rm theory}$ is obtained by these given $m_{\chi}$, $\delta$ and $F_{\rm{DM}}$ namely according to Eq.~\ref{Neevents}, while the $R_{1 e^{-}}^{\rm{obs}}$ is shown in Tab.\ref{SENSEI data}. The blank area between these colored dots represents the ratio $r>10$, which indicates that the theoretically calculated events $R_{1 e^{-}}^{\rm theory}$ is much larger than $R_{1 e^{-}}^{\rm{obs}}$. Whereas the ratio $r=0$ at the right bottom blank area implies $R_{1 e^{-}}^{\rm theory}=0$, which originates from two reasons.
For the up-scattering process with large $\delta$, the DM kinetic energy $E_{k}^{\chi}$ cannot sufficiently compensate for the large mass splitting $\delta$, leading to the up-scattering process being kinematically forbidden. Besides, the minimum velocity of incoming DM particles $v_{\rm{min}}$ for large $\delta$ will be larger than $v_{esc}+v_{E}$, causing the DM particles to escape the galaxy. Also, there exists the same constraint $(v_{\rm{min}} \leq v_{esc}+v_{E})$ for large $|\delta|$ down-scattering process. For a given DM mass $m_{\chi}$ and large mass splitting $|\delta|$ region, the small ratio $r=R_{1 e^{-}}^{\rm theory}/R_{1 e^{-}}^{\rm{obs}}$ indicates that the theoretically generated $R_{1 e^{-}}^{\rm{theory}}$ is much smaller than the experimentally observed $R_{1 e^{-}}^{\rm{obs}}$ for both up-scattering and down-scattering processes. This is because that for large mass splitting $|\delta|$ IDM-electron scattering process, the minimum velocity of incoming DM particles $v_{\rm{min}}$ is so large that the theoretically evaluated events $R_{1 e^{-}}^{\rm{theory}}$ are highly suppressed by the Maxwell-Boltzmann velocity distribution. Additionally, the less events $R_{1 e^{-}}^{\rm{theory}}$ generated by up-scattering process with large mass splitting $|\delta|$ also simultaneously arise from that more DM kinetic energy $E_{k}^{\chi}$ should be transformed to the large mass splitting $\delta$, remaining less kinetic energy to excite electrons. Note that the ratio $0<r \leq 1$ is allowed by the observed events from SENSEI experimental data with the reference cross section $\bar{\sigma}_{e}=10^{-37}$ cm$^{-2}$.

\begin{table}[ht]
    \centering
    \begin{tabular}{|c|c|c|c|c|}
    \hline
        $N_{e}$ & 1 & 2 & 3 & 4 \\
    \hline
    \hline
         Observed Events & 1311.7 & 5 & 0 & 0 \\
    \hline     
         90\%CL [g-day]$^{-1}$ & 525.2 & 4.449 & 0.255 & 0.233 \\
    \hline
    \end{tabular}
    \caption{The observed number of events and 90$\%$ CL [g-day]$^{-1}$ events from SENSEI experiment~\cite{SENSEI:2020dpa} data are shown in this table.}
    \label{SENSEI data}
\end{table}

\begin{figure}[ht]
\centering
\includegraphics[height=7cm,width=8cm]{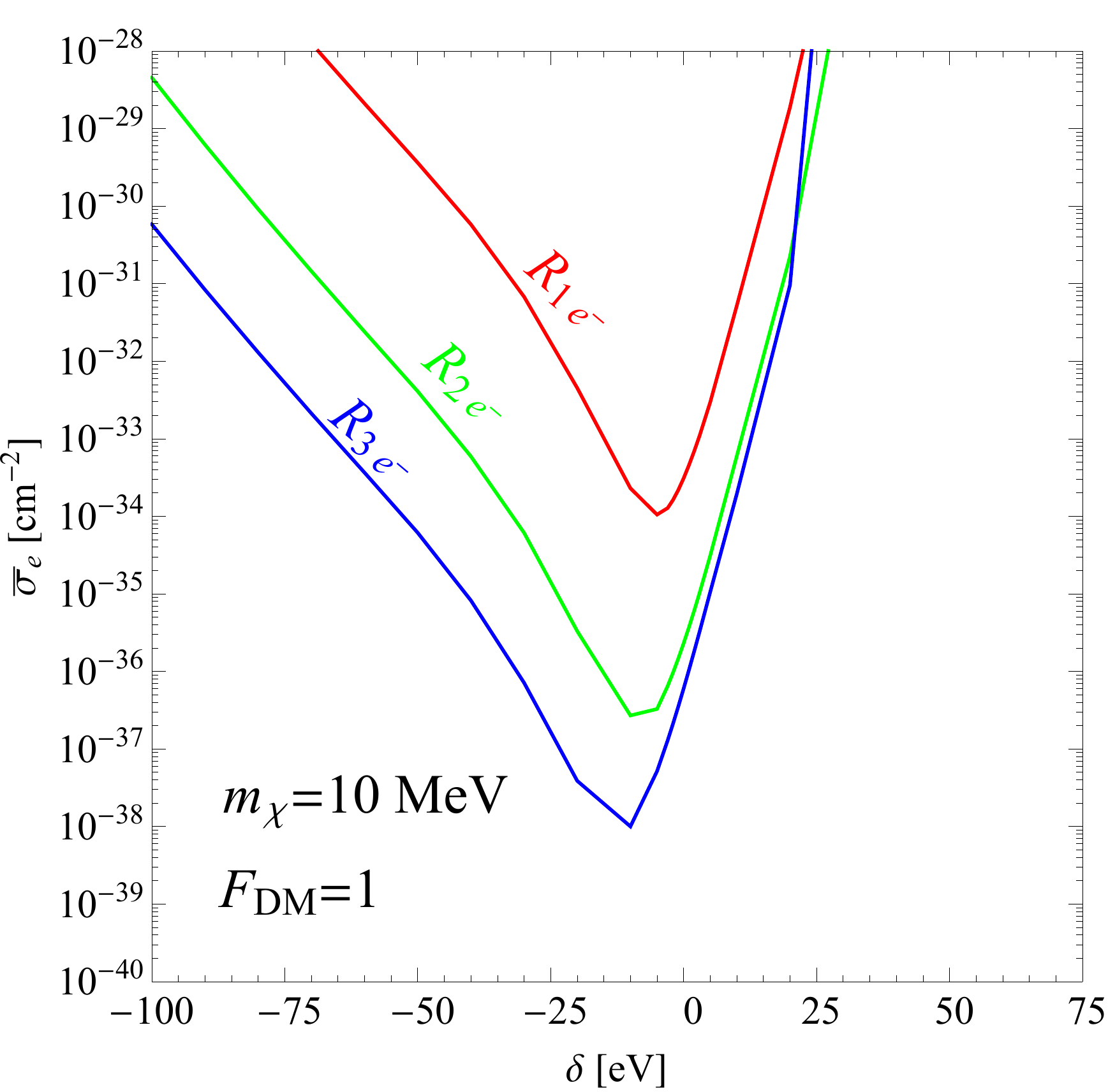}
\includegraphics[height=7cm,width=8cm]{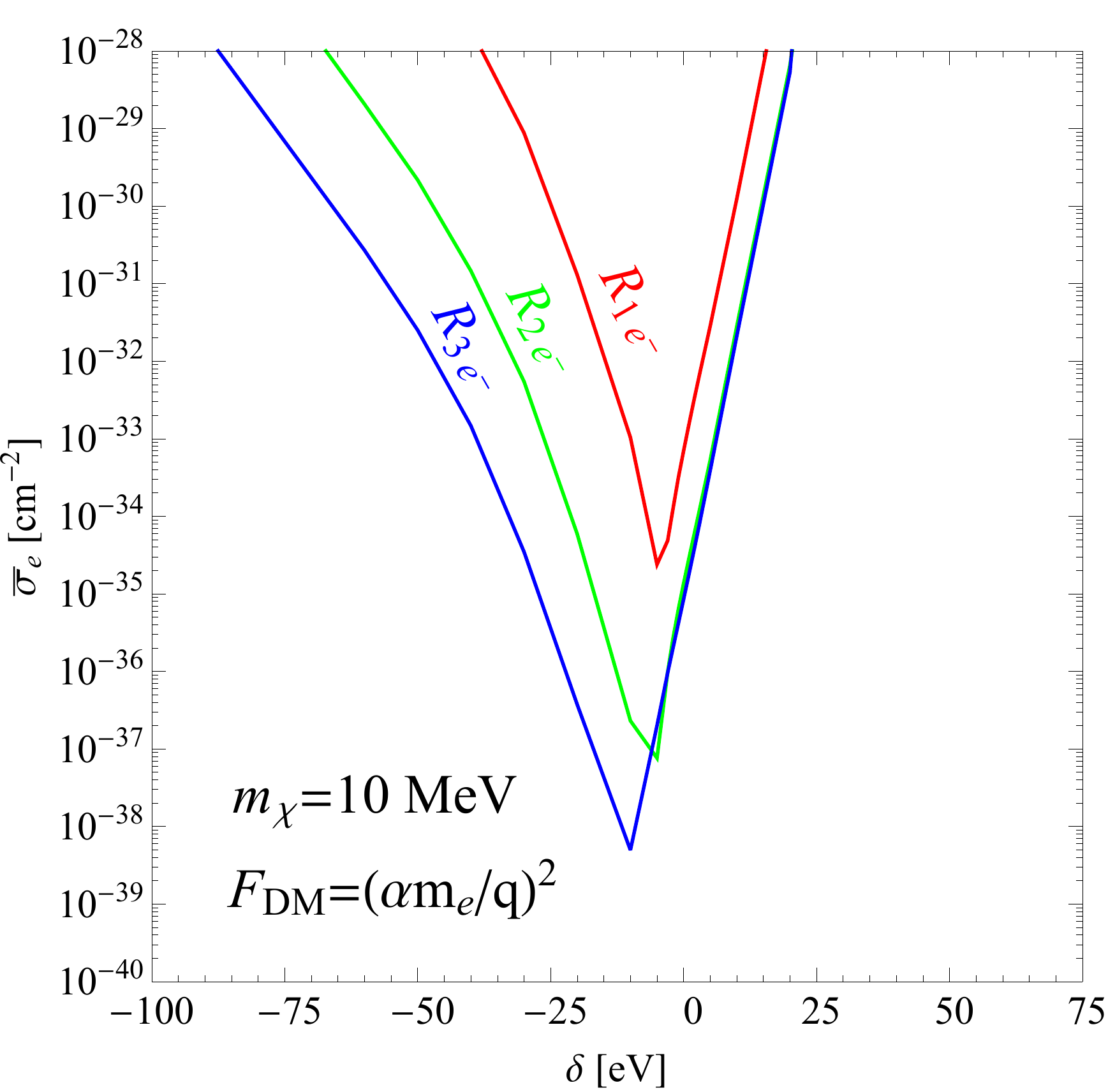}
\includegraphics[height=7cm,width=8cm]{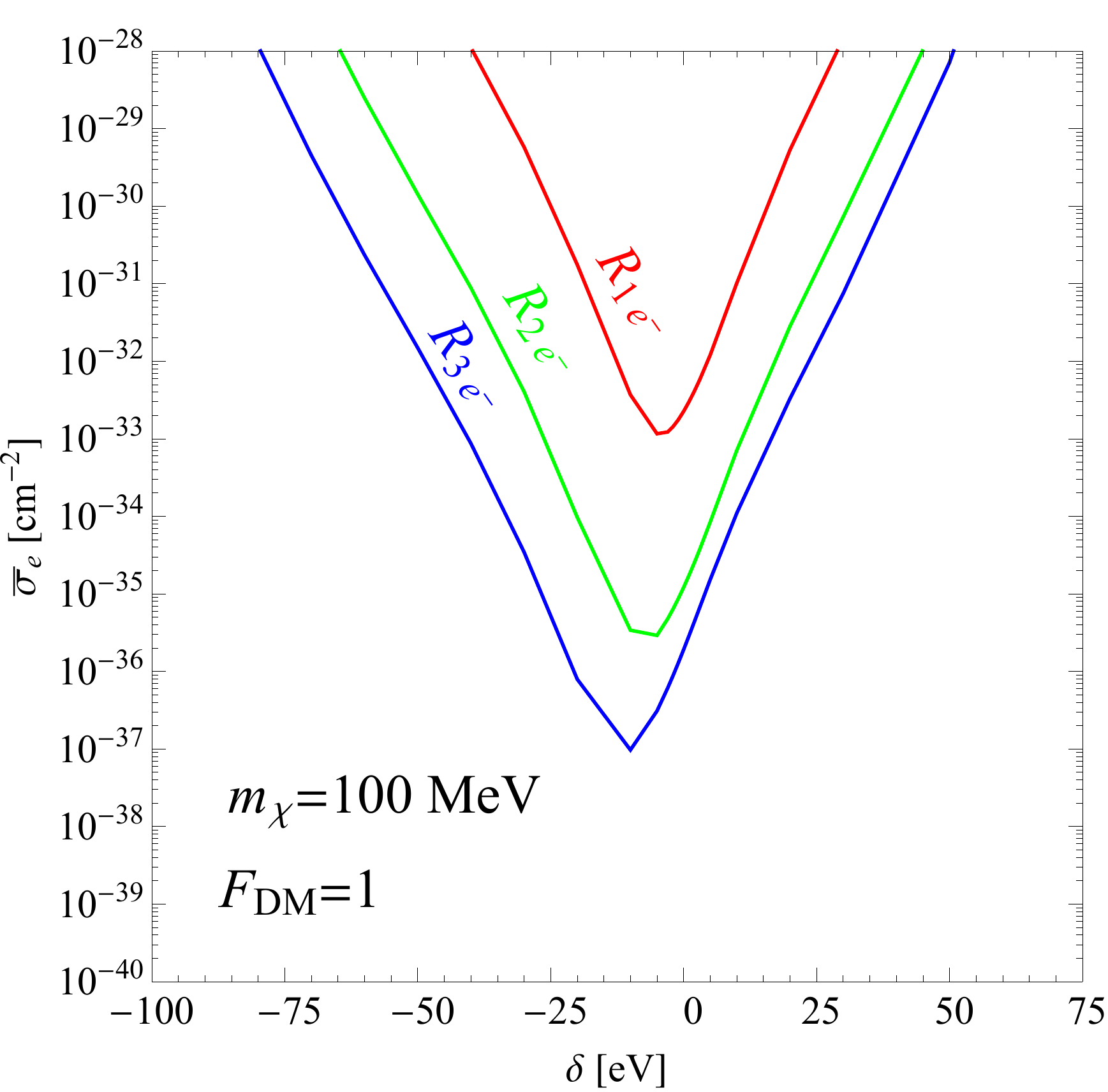}
\includegraphics[height=7cm,width=8cm]{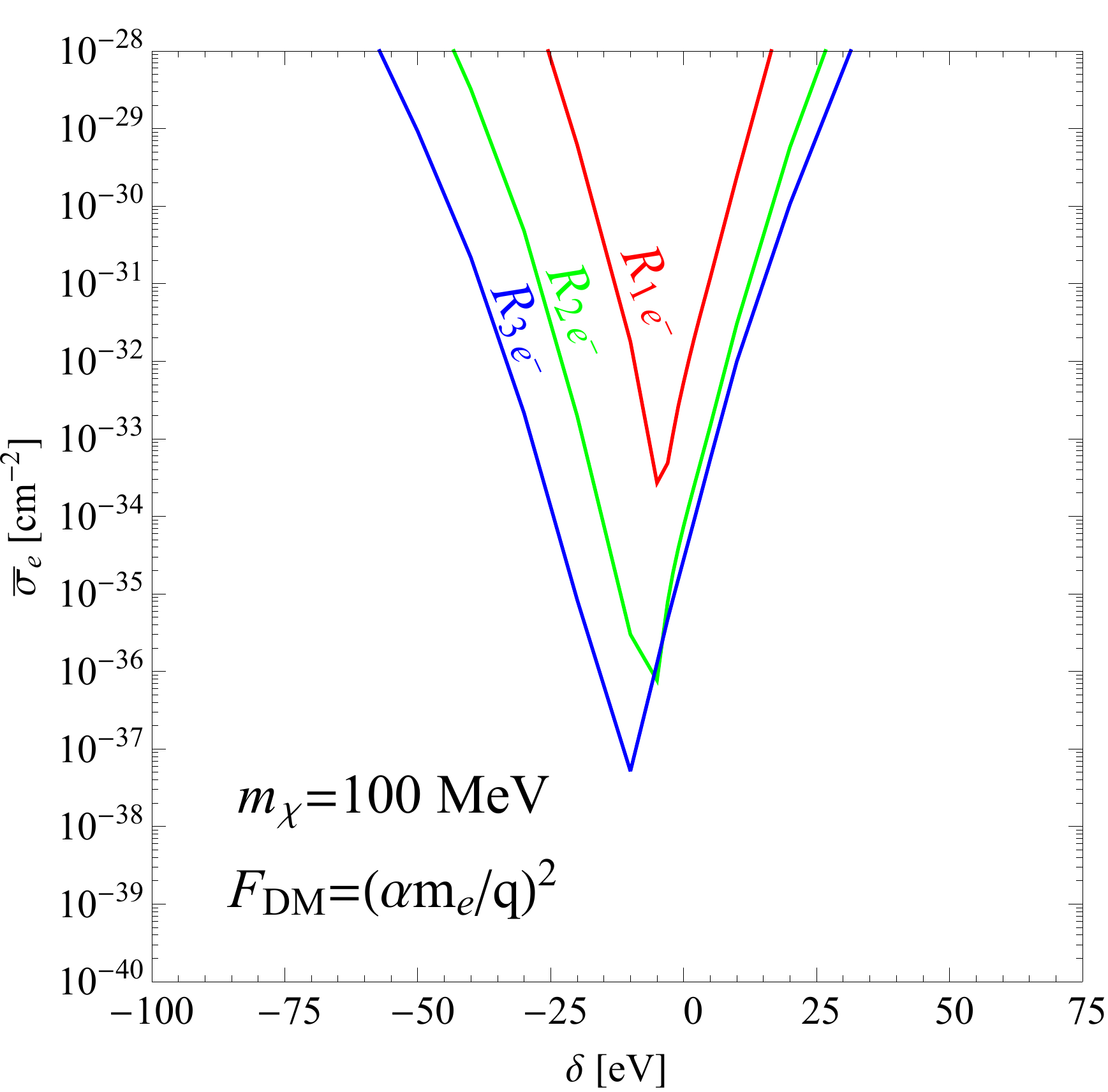}
\caption{The constraints on $\bar{\sigma}_{e}-\delta$ plane for two different dark matter masses $m_{\chi}=10$ MeV, $m_{\chi}=100$ MeV and DM form factor $F_{\rm{DM}}=1$, $F_{\rm{DM}}=(\alpha m_{e}/q)^2$ by using the data from SENSEI experiment.}
\label{sigmadelta}
\end{figure}

In Fig.~\ref{sigmadelta}, we utilize the different observable numbers of electrons to constrain the DM-electron scattering cross section $\bar{\sigma}_{e}$ and mass splitting $\delta$. The red, green and blue lines illustrate the limits from the different observed numbers of electrons (90$\%$CL [g-day]$^{-1}$) $R_{1e^{-}}^{\rm{obs}}(525.2), R_{2e^{-}}^{\rm{obs}}(4.449),R_{3e^{-}}^{\rm{obs}}(0.255)$ as shown in Tab.~\ref{SENSEI data}. With regard to the same DM mass $m_{\chi}$, form factor $F_{\rm{DM}}$ and mass splitting $\delta$, the more observed events $R_{Ne^{-}}^{\rm{obs}}$ will allow the larger DM-electron scattering cross section $\bar{\sigma}_{e}$. In other words, the more observed events $R_{Ne^{-}}^{\rm{obs}}$ put weaker limits on $\bar{\sigma}_{e}$.  Therefore, we can see that the least observed event $R_{3e^{-}}^{\rm{obs}}$ puts the most stringent limits on $\bar{\sigma}_{e}-\delta$ plane, while the most observed event $R_{1e^{-}}^{\rm{obs}}$ gives the weakest constraints in each picture in Fig.~\ref{sigmadelta}. As shown in Fig.~\ref{sigmadelta}, the constraints originating from up-scattering process will have a cutoff at the large $\delta$ where the DM kinetic energy $E_{k}^{\chi}$ cannot be enough transformed to the mass splitting or the corresponding minimum velocity of incoming DM particles $v_{\rm{min}}>v_{esc}+v_{E}$. In addition, we can see that the most stringent constraints occur at the mass splitting $\delta \sim \mathcal{O}(-10)$ eV, on either side of which the induced events are suppressed by the Maxwell-Boltzmann velocity distribution as mentioned before. One can notice that for up-scattering process $(\delta>0)$ with same DM form factor $F_{\rm{DM}}$ and mass splitting $\delta$, the heavy dark matter provides the stronger constraints on cross section $\bar{\sigma}_{e}$ while the light dark matter puts weaker limits on $\bar{\sigma}_{e}$. After overcoming the mass splitting $\delta$, the heavy dark matter has more kinetic energy left to generate more events $R_{Ne^{-}}^{\rm{theory}}$, so this will lead to the more stringent limits on $\bar{\sigma}_{e}$. Contrarily, with regard to down-scattering process $(\delta<0)$, the light dark matter gives stronger restrictions on $\bar{\sigma}_{e}$ while the heavy dark matter provides the weaker limits on $\bar{\sigma}_{e}$. For down-scattering process, due to the mass splitting contribution to extra energy to excite electrons, both the heavy and light dark matter have enough energy to excite all the electrons. However, the produced events in each energy bin are suppressed by the DM mass $m_{\chi}$, which gives rise to less events $R_{Ne^{-}}^{\rm{theory}}$ and weaker limits on $\bar{\sigma}_{e}$ for the heavy dark matter.
\begin{figure}[ht]
\centering
\includegraphics[height=7cm,width=8cm]{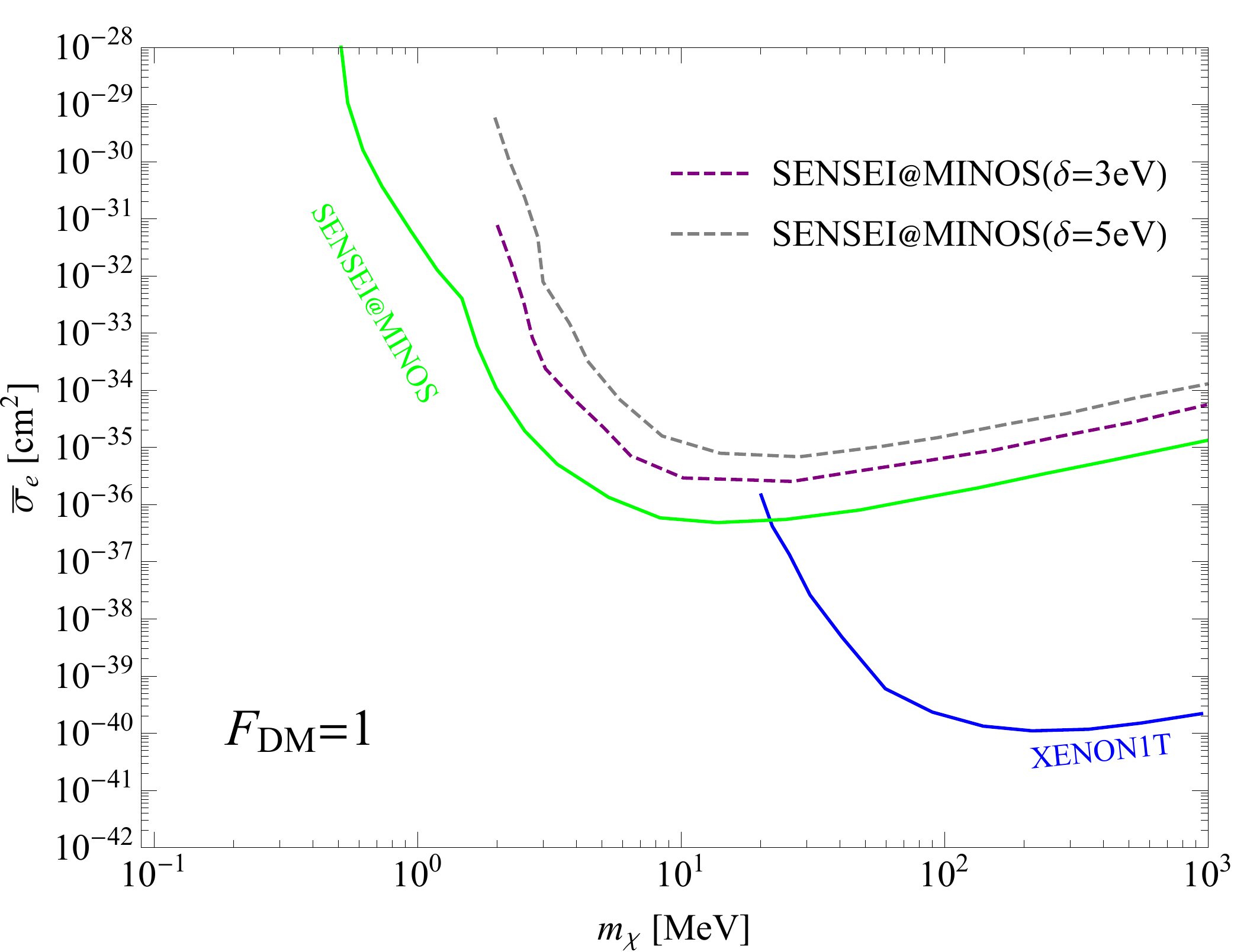}
\includegraphics[height=7cm,width=8cm]{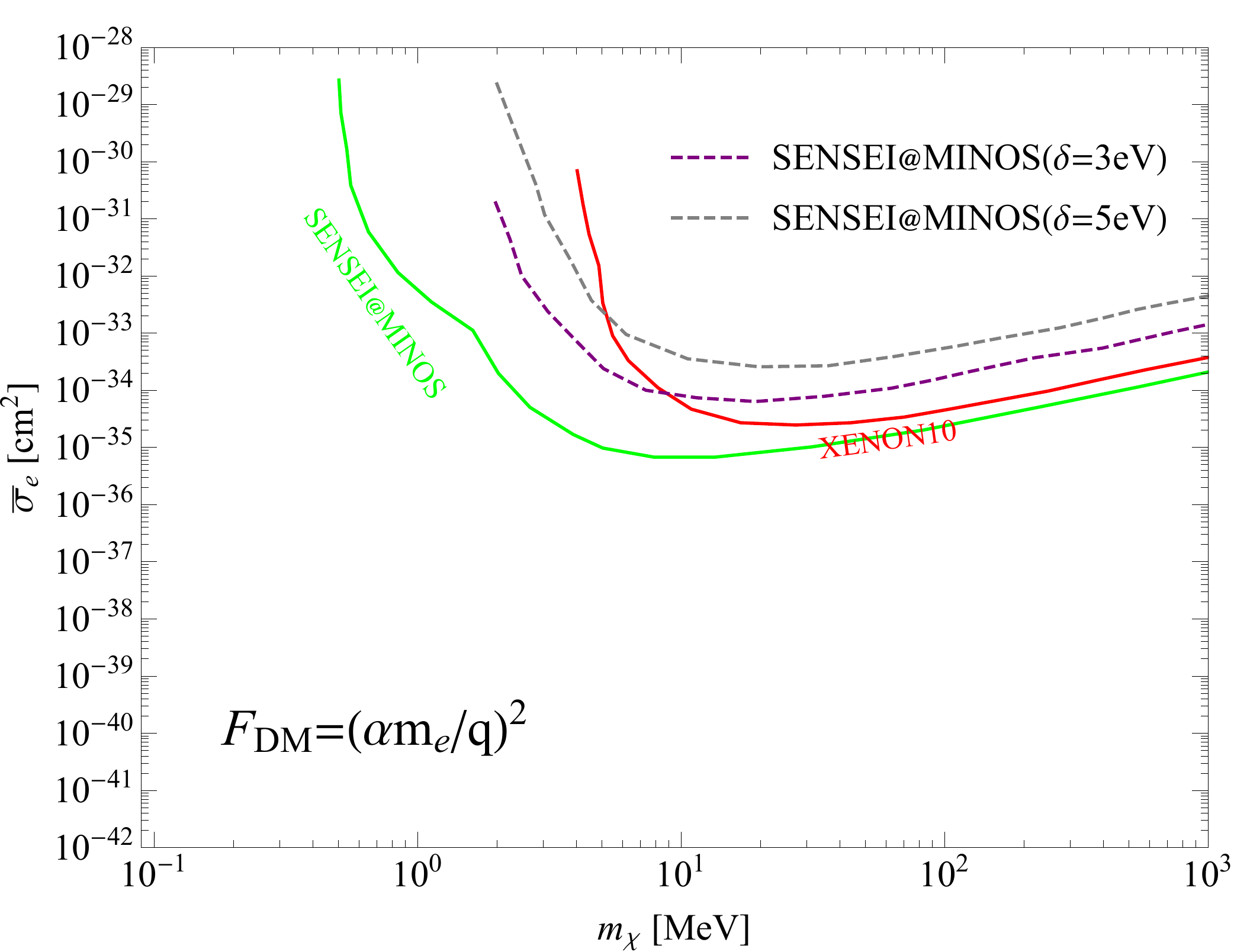}
\includegraphics[height=7cm,width=8cm]{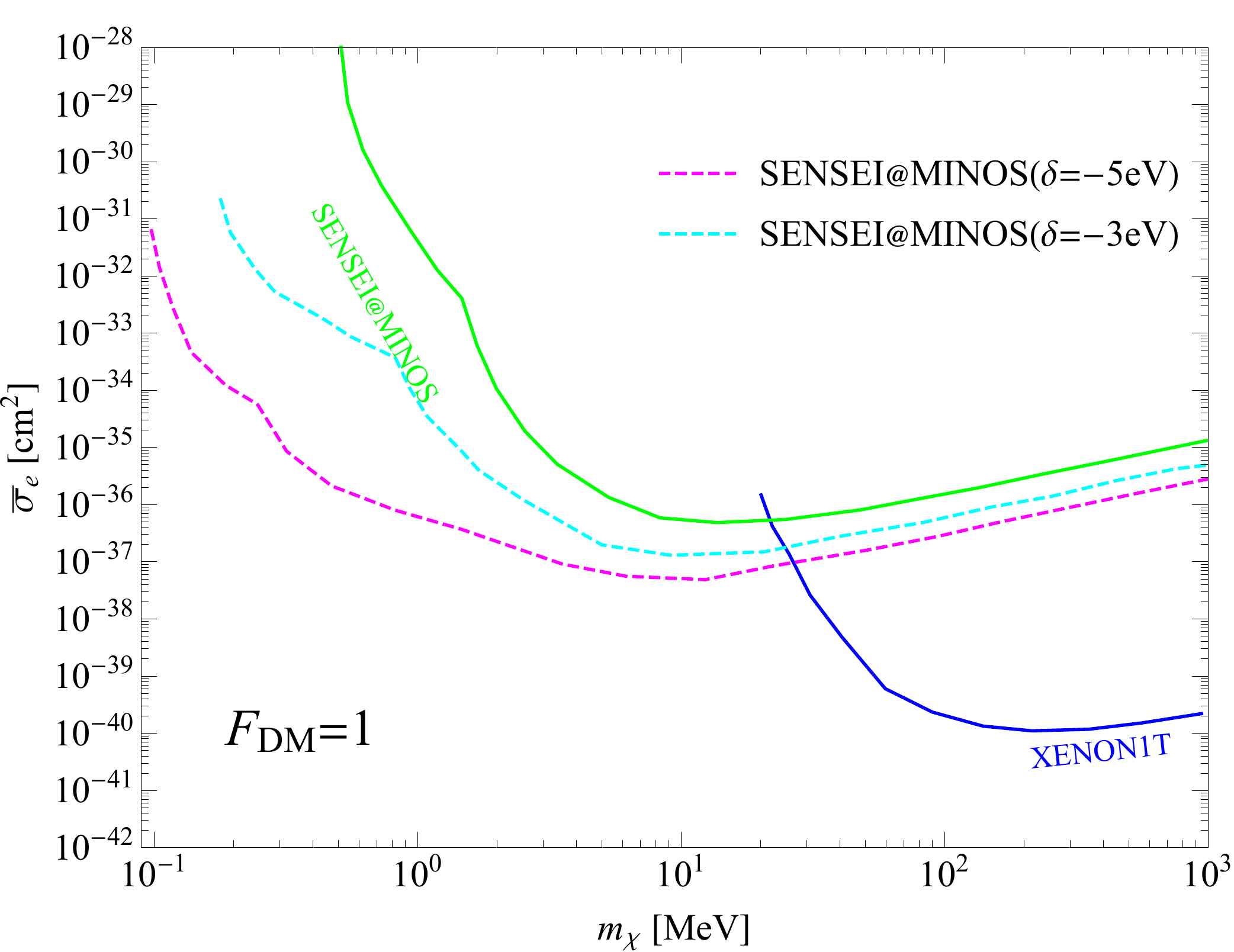}
\includegraphics[height=7cm,width=8cm]{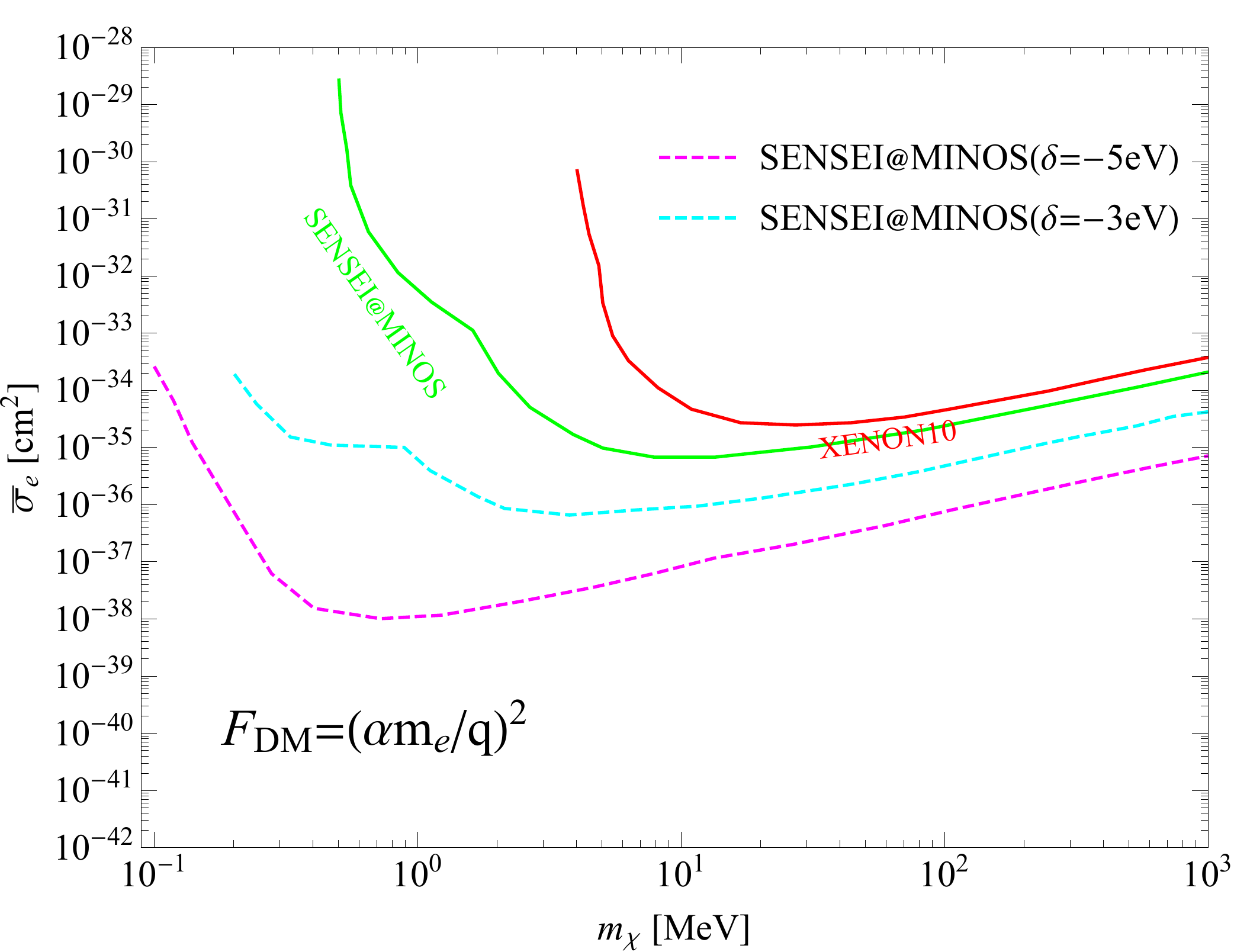}
\caption{The 90$\%$ CL constraints on DM-electron cross section $\bar{\sigma}_{e}$ versus DM mass $m_{\chi}$ for different mass splitting $\delta$ and two different DM form factors $F_{\rm{DM}}$ by using the latest published SENSEI@MINOS data. The main constraints on DM-electron cross section $\bar{\sigma}_{e}$ arise from experiments based on xenon targets: XENON10~\cite{Essig:2017kqs} and XENON1T~\cite{XENON:2019gfn}. The green solid line represents the limits on cross section $\bar{\sigma}_{e}$ from the released SENSEI experimental data while the blue(red) line illustrates the corresponding constraints on $\bar{\sigma}_{e}$ from XENON1T(XENON10) experimental data. }
\label{limits}
\end{figure}

In Fig.~\ref{limits}, we derive the limits on $m_{\chi}-\bar{\sigma}_{e}$ panel for different DM form factors $F_{\rm{DM}}$ and mass splitting $\delta$. The green solid line represents the constraint obtained by using the combined results of the different observed numbers of electrons from SENSEI experimental data (90$\%$CL [g-day]$^{-1}$) $R_{1e^{-}}^{\rm{obs}}(525.2), R_{2e^{-}}^{\rm{obs}}(4.449),R_{3e^{-}}^{\rm{obs}}(0.255),R_{4e^{-}}^{\rm{obs}}(0.253)$ as shown in Tab.~\ref{SENSEI data}. In each picture in Fig.\ref{limits}, the solid lines show the limits from the elastic DM-electron scattering, while the dashed lines represent the constraints from the IDM-electron scattering with different mass splitting $\delta$ in the inelastic dark matter model. The two upper pictures indicate the up-scattering process, whereas the two bottom pictures represent the down-scattering process. As we can see in Fig.~\ref{limits}, compared with the elastic DM-electron scattering process(the green solid line), the down-scattering process $\delta<0$ gets the stronger constraints on $\bar{\sigma}_{e}$ due to its resulting in more events $R_{Ne^{-}}^{\rm{theory}}$ as mentioned before. Whereas the up-scattering process provides the relatively weaker restrictions on $\bar{\sigma}_{e}$ because the part of the DM kinetic energy $E_{k}^{\chi}$ will be converted to mass splitting $\delta$, rather than absolutely transformed to the electron deposited energy $\Delta E_{e}$. This will generate the less events $R_{Ne^{-}}^{\rm{theory}}$. Besides, the constraints become weaker with the mass splitting $\delta$ increasing for up-scattering process since more DM kinetic energy $E_{k}^{\chi}$ has to be converted the mass splitting $\delta$, which leads to the less observable events. However, the limits are more stringent with the mass splitting $|\delta|$ increasing for down-scattering process. This is because that large mass splitting $|\delta|$ implies that more extra energy induces the more observable events. But, the resulting events are also highly suppressed by the Maxwell-Boltzmann velocity distribution since large mass splitting $|\delta|$ also indicates large $v_{\rm min}$ as mentioned before. The number of generated events $R_{Ne^{-}}^{\rm{theory}}$ depends on the competition between the large $\delta$ enhancement and Maxwell-Boltzmann velocity distribution suppression.  Therefore, we take the mass splitting $\delta=-3, -5$ eV for down-scattering as a benchmark point from Fig~\ref{sigmadelta}. Besides, XENON1T excess can be accounted for by the inelastic dark matter with the mass splitting $\delta \sim 2-3$ keV located at the peak of the electron recoil energy spectrum excess. The down-scattering process with $|\delta| \sim 2-3$ keV, on the other hand, gives rise to less events because of the Maxwell-Boltzmann velocity distribution suppression. Therefore, the constraints from the down-scattering process with $|\delta| \sim 2-3$  keV are weaker than those shown in Fig.~\ref{limits}.\\
\section{Conclusion}
\label{sec4}

Given the current status of searching for WIMP dark matter, the searches for the light dark matter have attracted a great amount of attention. Light dark matter can reside in some well-motivated models such as the inelastic dark matter model. In this work, we have studied the IDM-electron scattering in silicon semiconductors due to their lower binding energy. Furthermore, the SENSEI experiment with the ultralow-noise silicon Skipper-CCD has given strong limits on cross section $\bar{\sigma}_{e}$. We utilize the latest released data from SENSEI experiment to constrain the cross section $\bar{\sigma}_{e}$ in the inelastic dark matter model. With regard to IDM-electron scattering, the SENSEI experiment gives the stronger(weaker) constraints on cross section $\bar{\sigma}_{e}$ for down-scattering (up-scattering) process. Especially, the SENSEI experiment can detect the DM mass down to 0.1 MeV for down-scattering process with the mass splitting $\delta \sim -5$ eV.

\section{Acknowledgements}
\label{sec5}
We thank T.-T. Yu for helpful discussions. This work is supported by the National Natural Science Foundation of China (NNSFC) under grant No. 11805161, by Natural Science Foundation of Shandong Province under the grants ZR2018QA007.

\section{Appendix: Numerical Calculation of the Crystal Form Factor}
\label{sec6}

The numerical crystal form factor average over bins of equal width in $q$ and $E_{e}$ and is expressed by   
\begin{equation}
    \left|f_{c}(q_{i},\Delta E_{j})\right|^2 \equiv \int^{q_{i}+\frac{1}{2}\delta q}_{q_{i}-\frac{1}{2}\delta q} \frac{dq^{'}}{\delta q} \int^{\Delta E_{j}+\frac{1}{2}\delta E}_{\Delta E_{j}-\frac{1}{2}\delta E} \frac{d \Delta E^{'}}{\delta E} \left|f_{c}(q^{'},\Delta E^{'})\right|^2,
\end{equation}
where $q_{i}(\Delta E_{j})$ is the central value of $i$-th $q$ bin($j$-th energy bin). The range of $\Delta E_{e}$ is from 0.1 eV to 50 eV with 500 bins at intervals $\delta E_{e}=0.1$ eV, while the range of $q$ is from 0.02 $\alpha m_{e}$ to 18 $\alpha m_{e}$ with 900 bins at intervals $\delta q=0.02 \alpha m_{e}$. In addition to binning in $q$ and $\Delta E_{e}$, the reciprocal lattice vectors $\bf{G}$ follows the cutoff requirement
\begin{equation}
    \frac{\left|\bf{k}+\bf{G}\right|^2}{2 m_{e}} \leq E_{\rm{cut}},
\end{equation}
resulting in a fundamental cutoff on transfer momentum $q \leq \sqrt{2 m_{e} E_{\rm{cut}}}$, where $E_{\rm{cut}}$ is the plane-wave energy cutoff. We take the same $E_{\rm{cut}}=70$ Ry value as mentioned in Refs~\cite{Essig:2015cda}. Furthermore, the numerical calculation also requires replacing the $k$-integral with a discretization in $\bf{k}$.
\begin{eqnarray}
   \left|f_{c}^{N}(q_{i},\Delta E_{j})\right|^2 &=& \frac{2 \pi^2 (\alpha m_{e}^2 V_{\rm{cell}})^{-1}}{E_{j}} \sum_{n,n^{'}} \sum_{\bf{k},\bf{k}^{'}}\sum_{\bf{G}^{'}} \frac{q_{i}}{\delta q} \frac{E_{j}}{\delta E} \frac{\omega_{\bf{k}}}{2} \frac{\omega_{\bf{k}^{'}}}{2} \left|f_{n{\bf{k}}\rightarrow n^{'}{\bf{k}^{'},\bf{G}^{'}}}\right|^2\nonumber \\
   &\times& \Theta\left(1-\frac{\left|E_{n^{'}, \bf{k}^{'}}-E_{n,\bf{k}}-\Delta E_{j}\right|}{\frac{1}{2}\delta E}\right) \Theta\left(1-\frac{\left||\bf{k}^{'}-\bf{k}+{\bf{G}^{'}}|-q_{i}\right|}{\frac{1}{2}\delta q}\right),
\end{eqnarray}
where the 243 representative $k$-points with the corresponding weightings $\omega_{\bf{k}}$ are used in the sum of $\bf{k}$ and the weightings $\omega_{\bf{k}}$ satisfy the condition $\sum_{k} \omega_{\bf{k}}=2$. 

\bibliography{Refs}

\begin{thebibliography}{92}
\expandafter\ifx\csname natexlab\endcsname\relax\def\natexlab#1{#1}\fi
\expandafter\ifx\csname bibnamefont\endcsname\relax
  \def\bibnamefont#1{#1}\fi
\expandafter\ifx\csname bibfnamefont\endcsname\relax
  \def\bibfnamefont#1{#1}\fi
\expandafter\ifx\csname citenamefont\endcsname\relax
  \def\citenamefont#1{#1}\fi
\expandafter\ifx\csname url\endcsname\relax
  \def\url#1{\texttt{#1}}\fi
\expandafter\ifx\csname urlprefix\endcsname\relax\def\urlprefix{URL }\fi
\providecommand{\bibinfo}[2]{#2}
\providecommand{\eprint}[2][]{\url{#2}}

\bibitem[{\citenamefont{Trimble}(1987)}]{Trimble:1987ee}
\bibinfo{author}{\bibfnamefont{V.}~\bibnamefont{Trimble}},
  \bibinfo{journal}{Ann. Rev. Astron. Astrophys.}
  \textbf{\bibinfo{volume}{25}}, \bibinfo{pages}{425} (\bibinfo{year}{1987}).

\bibitem[{\citenamefont{Schumann}(2019)}]{Schumann:2019eaa}
\bibinfo{author}{\bibfnamefont{M.}~\bibnamefont{Schumann}},
  \bibinfo{journal}{J. Phys. G} \textbf{\bibinfo{volume}{46}},
  \bibinfo{pages}{103003} (\bibinfo{year}{2019}), \eprint{1903.03026}.

\bibitem[{\citenamefont{Kouvaris and Pradler}(2017)}]{Kouvaris:2016afs}
\bibinfo{author}{\bibfnamefont{C.}~\bibnamefont{Kouvaris}} \bibnamefont{and}
  \bibinfo{author}{\bibfnamefont{J.}~\bibnamefont{Pradler}},
  \bibinfo{journal}{Phys. Rev. Lett.} \textbf{\bibinfo{volume}{118}},
  \bibinfo{pages}{031803} (\bibinfo{year}{2017}), \eprint{1607.01789}.

\bibitem[{\citenamefont{Grilli~di Cortona et~al.}(2020)\citenamefont{Grilli~di
  Cortona, Messina, and Piacentini}}]{GrillidiCortona:2020owp}
\bibinfo{author}{\bibfnamefont{G.}~\bibnamefont{Grilli~di Cortona}},
  \bibinfo{author}{\bibfnamefont{A.}~\bibnamefont{Messina}}, \bibnamefont{and}
  \bibinfo{author}{\bibfnamefont{S.}~\bibnamefont{Piacentini}},
  \bibinfo{journal}{JHEP} \textbf{\bibinfo{volume}{11}}, \bibinfo{pages}{034}
  (\bibinfo{year}{2020}), \eprint{2006.02453}.

\bibitem[{\citenamefont{Ibe et~al.}(2018)\citenamefont{Ibe, Nakano, Shoji, and
  Suzuki}}]{Ibe:2017yqa}
\bibinfo{author}{\bibfnamefont{M.}~\bibnamefont{Ibe}},
  \bibinfo{author}{\bibfnamefont{W.}~\bibnamefont{Nakano}},
  \bibinfo{author}{\bibfnamefont{Y.}~\bibnamefont{Shoji}}, \bibnamefont{and}
  \bibinfo{author}{\bibfnamefont{K.}~\bibnamefont{Suzuki}},
  \bibinfo{journal}{JHEP} \textbf{\bibinfo{volume}{03}}, \bibinfo{pages}{194}
  (\bibinfo{year}{2018}), \eprint{1707.07258}.

\bibitem[{\citenamefont{Dolan et~al.}(2018)\citenamefont{Dolan, Kahlhoefer, and
  McCabe}}]{Dolan:2017xbu}
\bibinfo{author}{\bibfnamefont{M.~J.} \bibnamefont{Dolan}},
  \bibinfo{author}{\bibfnamefont{F.}~\bibnamefont{Kahlhoefer}},
  \bibnamefont{and} \bibinfo{author}{\bibfnamefont{C.}~\bibnamefont{McCabe}},
  \bibinfo{journal}{Phys. Rev. Lett.} \textbf{\bibinfo{volume}{121}},
  \bibinfo{pages}{101801} (\bibinfo{year}{2018}), \eprint{1711.09906}.

\bibitem[{\citenamefont{Bell et~al.}(2020)\citenamefont{Bell, Dent, Newstead,
  Sabharwal, and Weiler}}]{Bell:2019egg}
\bibinfo{author}{\bibfnamefont{N.~F.} \bibnamefont{Bell}},
  \bibinfo{author}{\bibfnamefont{J.~B.} \bibnamefont{Dent}},
  \bibinfo{author}{\bibfnamefont{J.~L.} \bibnamefont{Newstead}},
  \bibinfo{author}{\bibfnamefont{S.}~\bibnamefont{Sabharwal}},
  \bibnamefont{and} \bibinfo{author}{\bibfnamefont{T.~J.}
  \bibnamefont{Weiler}}, \bibinfo{journal}{Phys. Rev. D}
  \textbf{\bibinfo{volume}{101}}, \bibinfo{pages}{015012}
  (\bibinfo{year}{2020}), \eprint{1905.00046}.

\bibitem[{\citenamefont{Essig et~al.}(2020)\citenamefont{Essig, Pradler,
  Sholapurkar, and Yu}}]{Essig:2019xkx}
\bibinfo{author}{\bibfnamefont{R.}~\bibnamefont{Essig}},
  \bibinfo{author}{\bibfnamefont{J.}~\bibnamefont{Pradler}},
  \bibinfo{author}{\bibfnamefont{M.}~\bibnamefont{Sholapurkar}},
  \bibnamefont{and} \bibinfo{author}{\bibfnamefont{T.-T.} \bibnamefont{Yu}},
  \bibinfo{journal}{Phys. Rev. Lett.} \textbf{\bibinfo{volume}{124}},
  \bibinfo{pages}{021801} (\bibinfo{year}{2020}), \eprint{1908.10881}.

\bibitem[{\citenamefont{Knapen et~al.}(2021)\citenamefont{Knapen, Kozaczuk, and
  Lin}}]{Knapen:2020aky}
\bibinfo{author}{\bibfnamefont{S.}~\bibnamefont{Knapen}},
  \bibinfo{author}{\bibfnamefont{J.}~\bibnamefont{Kozaczuk}}, \bibnamefont{and}
  \bibinfo{author}{\bibfnamefont{T.}~\bibnamefont{Lin}},
  \bibinfo{journal}{Phys. Rev. Lett.} \textbf{\bibinfo{volume}{127}},
  \bibinfo{pages}{081805} (\bibinfo{year}{2021}), \eprint{2011.09496}.

\bibitem[{\citenamefont{Flambaum et~al.}(2020)\citenamefont{Flambaum, Su, Wu,
  and Zhu}}]{Flambaum:2020xxo}
\bibinfo{author}{\bibfnamefont{V.~V.} \bibnamefont{Flambaum}},
  \bibinfo{author}{\bibfnamefont{L.}~\bibnamefont{Su}},
  \bibinfo{author}{\bibfnamefont{L.}~\bibnamefont{Wu}}, \bibnamefont{and}
  \bibinfo{author}{\bibfnamefont{B.}~\bibnamefont{Zhu}} (\bibinfo{year}{2020}),
  \eprint{2012.09751}.

\bibitem[{\citenamefont{Bell et~al.}(2021{\natexlab{a}})\citenamefont{Bell,
  Dent, Dutta, Ghosh, Kumar, and Newstead}}]{Bell:2021zkr}
\bibinfo{author}{\bibfnamefont{N.~F.} \bibnamefont{Bell}},
  \bibinfo{author}{\bibfnamefont{J.~B.} \bibnamefont{Dent}},
  \bibinfo{author}{\bibfnamefont{B.}~\bibnamefont{Dutta}},
  \bibinfo{author}{\bibfnamefont{S.}~\bibnamefont{Ghosh}},
  \bibinfo{author}{\bibfnamefont{J.}~\bibnamefont{Kumar}}, \bibnamefont{and}
  \bibinfo{author}{\bibfnamefont{J.~L.} \bibnamefont{Newstead}},
  \bibinfo{journal}{Phys. Rev. D} \textbf{\bibinfo{volume}{104}},
  \bibinfo{pages}{076013} (\bibinfo{year}{2021}{\natexlab{a}}),
  \eprint{2103.05890}.

\bibitem[{\citenamefont{Wang et~al.}(2021)\citenamefont{Wang, Wu, Wu, and
  Zhu}}]{Wang:2021oha}
\bibinfo{author}{\bibfnamefont{W.}~\bibnamefont{Wang}},
  \bibinfo{author}{\bibfnamefont{K.-Y.} \bibnamefont{Wu}},
  \bibinfo{author}{\bibfnamefont{L.}~\bibnamefont{Wu}}, \bibnamefont{and}
  \bibinfo{author}{\bibfnamefont{B.}~\bibnamefont{Zhu}} (\bibinfo{year}{2021}),
  \eprint{2112.06492}.

\bibitem[{\citenamefont{Essig et~al.}(2012{\natexlab{a}})\citenamefont{Essig,
  Mardon, and Volansky}}]{Essig:2011nj}
\bibinfo{author}{\bibfnamefont{R.}~\bibnamefont{Essig}},
  \bibinfo{author}{\bibfnamefont{J.}~\bibnamefont{Mardon}}, \bibnamefont{and}
  \bibinfo{author}{\bibfnamefont{T.}~\bibnamefont{Volansky}},
  \bibinfo{journal}{Phys. Rev. D} \textbf{\bibinfo{volume}{85}},
  \bibinfo{pages}{076007} (\bibinfo{year}{2012}{\natexlab{a}}),
  \eprint{1108.5383}.

\bibitem[{\citenamefont{Bloch et~al.}(2021)\citenamefont{Bloch, Caputo, Essig,
  Redigolo, Sholapurkar, and Volansky}}]{Bloch:2020uzh}
\bibinfo{author}{\bibfnamefont{I.~M.} \bibnamefont{Bloch}},
  \bibinfo{author}{\bibfnamefont{A.}~\bibnamefont{Caputo}},
  \bibinfo{author}{\bibfnamefont{R.}~\bibnamefont{Essig}},
  \bibinfo{author}{\bibfnamefont{D.}~\bibnamefont{Redigolo}},
  \bibinfo{author}{\bibfnamefont{M.}~\bibnamefont{Sholapurkar}},
  \bibnamefont{and} \bibinfo{author}{\bibfnamefont{T.}~\bibnamefont{Volansky}},
  \bibinfo{journal}{JHEP} \textbf{\bibinfo{volume}{01}}, \bibinfo{pages}{178}
  (\bibinfo{year}{2021}), \eprint{2006.14521}.

\bibitem[{\citenamefont{Gao et~al.}(2020)\citenamefont{Gao, Liu, Wang, Wang,
  Xue, and Zhong}}]{Gao:2020wer}
\bibinfo{author}{\bibfnamefont{C.}~\bibnamefont{Gao}},
  \bibinfo{author}{\bibfnamefont{J.}~\bibnamefont{Liu}},
  \bibinfo{author}{\bibfnamefont{L.-T.} \bibnamefont{Wang}},
  \bibinfo{author}{\bibfnamefont{X.-P.} \bibnamefont{Wang}},
  \bibinfo{author}{\bibfnamefont{W.}~\bibnamefont{Xue}}, \bibnamefont{and}
  \bibinfo{author}{\bibfnamefont{Y.-M.} \bibnamefont{Zhong}},
  \bibinfo{journal}{Phys. Rev. Lett.} \textbf{\bibinfo{volume}{125}},
  \bibinfo{pages}{131806} (\bibinfo{year}{2020}), \eprint{2006.14598}.

\bibitem[{\citenamefont{Ge et~al.}(2021)\citenamefont{Ge, Liu, Yuan, and
  Zhou}}]{Ge:2020yuf}
\bibinfo{author}{\bibfnamefont{S.-F.} \bibnamefont{Ge}},
  \bibinfo{author}{\bibfnamefont{J.}~\bibnamefont{Liu}},
  \bibinfo{author}{\bibfnamefont{Q.}~\bibnamefont{Yuan}}, \bibnamefont{and}
  \bibinfo{author}{\bibfnamefont{N.}~\bibnamefont{Zhou}},
  \bibinfo{journal}{Phys. Rev. Lett.} \textbf{\bibinfo{volume}{126}},
  \bibinfo{pages}{091804} (\bibinfo{year}{2021}), \eprint{2005.09480}.

\bibitem[{\citenamefont{Athron et~al.}(2021)}]{Athron:2020maw}
\bibinfo{author}{\bibfnamefont{P.}~\bibnamefont{Athron}} \bibnamefont{et~al.},
  \bibinfo{journal}{JHEP} \textbf{\bibinfo{volume}{05}}, \bibinfo{pages}{159}
  (\bibinfo{year}{2021}), \eprint{2007.05517}.

\bibitem[{\citenamefont{Su et~al.}(2020)\citenamefont{Su, Wang, Wu, Yang, and
  Zhu}}]{Su:2020zny}
\bibinfo{author}{\bibfnamefont{L.}~\bibnamefont{Su}},
  \bibinfo{author}{\bibfnamefont{W.}~\bibnamefont{Wang}},
  \bibinfo{author}{\bibfnamefont{L.}~\bibnamefont{Wu}},
  \bibinfo{author}{\bibfnamefont{J.~M.} \bibnamefont{Yang}}, \bibnamefont{and}
  \bibinfo{author}{\bibfnamefont{B.}~\bibnamefont{Zhu}},
  \bibinfo{journal}{Phys. Rev. D} \textbf{\bibinfo{volume}{102}},
  \bibinfo{pages}{115028} (\bibinfo{year}{2020}), \eprint{2006.11837}.

\bibitem[{\citenamefont{Cao et~al.}(2021)\citenamefont{Cao, Ding, and
  Xiang}}]{Cao:2020bwd}
\bibinfo{author}{\bibfnamefont{Q.-H.} \bibnamefont{Cao}},
  \bibinfo{author}{\bibfnamefont{R.}~\bibnamefont{Ding}}, \bibnamefont{and}
  \bibinfo{author}{\bibfnamefont{Q.-F.} \bibnamefont{Xiang}},
  \bibinfo{journal}{Chin. Phys. C} \textbf{\bibinfo{volume}{45}},
  \bibinfo{pages}{045002} (\bibinfo{year}{2021}), \eprint{2006.12767}.

\bibitem[{\citenamefont{An et~al.}(2020)\citenamefont{An, Pospelov, Pradler,
  and Ritz}}]{An:2020bxd}
\bibinfo{author}{\bibfnamefont{H.}~\bibnamefont{An}},
  \bibinfo{author}{\bibfnamefont{M.}~\bibnamefont{Pospelov}},
  \bibinfo{author}{\bibfnamefont{J.}~\bibnamefont{Pradler}}, \bibnamefont{and}
  \bibinfo{author}{\bibfnamefont{A.}~\bibnamefont{Ritz}},
  \bibinfo{journal}{Phys. Rev. D} \textbf{\bibinfo{volume}{102}},
  \bibinfo{pages}{115022} (\bibinfo{year}{2020}), \eprint{2006.13929}.

\bibitem[{\citenamefont{Zu et~al.}(2021)\citenamefont{Zu, Foot, Fan, and
  Feng}}]{Zu:2020bsx}
\bibinfo{author}{\bibfnamefont{L.}~\bibnamefont{Zu}},
  \bibinfo{author}{\bibfnamefont{R.}~\bibnamefont{Foot}},
  \bibinfo{author}{\bibfnamefont{Y.-Z.} \bibnamefont{Fan}}, \bibnamefont{and}
  \bibinfo{author}{\bibfnamefont{L.}~\bibnamefont{Feng}},
  \bibinfo{journal}{JCAP} \textbf{\bibinfo{volume}{01}}, \bibinfo{pages}{070}
  (\bibinfo{year}{2021}), \eprint{2007.15191}.

\bibitem[{\citenamefont{Guo et~al.}(2020)\citenamefont{Guo, Tsai, Wu, and
  Yuan}}]{Guo:2020oum}
\bibinfo{author}{\bibfnamefont{G.}~\bibnamefont{Guo}},
  \bibinfo{author}{\bibfnamefont{Y.-L.~S.} \bibnamefont{Tsai}},
  \bibinfo{author}{\bibfnamefont{M.-R.} \bibnamefont{Wu}}, \bibnamefont{and}
  \bibinfo{author}{\bibfnamefont{Q.}~\bibnamefont{Yuan}},
  \bibinfo{journal}{Phys. Rev. D} \textbf{\bibinfo{volume}{102}},
  \bibinfo{pages}{103004} (\bibinfo{year}{2020}), \eprint{2008.12137}.

\bibitem[{\citenamefont{Du et~al.}(2021)\citenamefont{Du, Liang, Liu, Tran, and
  Xue}}]{Du:2020ybt}
\bibinfo{author}{\bibfnamefont{M.}~\bibnamefont{Du}},
  \bibinfo{author}{\bibfnamefont{J.}~\bibnamefont{Liang}},
  \bibinfo{author}{\bibfnamefont{Z.}~\bibnamefont{Liu}},
  \bibinfo{author}{\bibfnamefont{V.~Q.} \bibnamefont{Tran}}, \bibnamefont{and}
  \bibinfo{author}{\bibfnamefont{Y.}~\bibnamefont{Xue}},
  \bibinfo{journal}{Chin. Phys. C} \textbf{\bibinfo{volume}{45}},
  \bibinfo{pages}{013114} (\bibinfo{year}{2021}), \eprint{2006.11949}.

\bibitem[{\citenamefont{Chao et~al.}(2021)\citenamefont{Chao, Jin, and
  Peng}}]{Chao:2021liw}
\bibinfo{author}{\bibfnamefont{W.}~\bibnamefont{Chao}},
  \bibinfo{author}{\bibfnamefont{M.}~\bibnamefont{Jin}}, \bibnamefont{and}
  \bibinfo{author}{\bibfnamefont{Y.-Q.} \bibnamefont{Peng}}
  (\bibinfo{year}{2021}), \eprint{2109.14944}.

\bibitem[{\citenamefont{Chen et~al.}(2021)\citenamefont{Chen, Fornal, Sandick,
  Shu, Xue, Zhao, and Zong}}]{Chen:2021ifo}
\bibinfo{author}{\bibfnamefont{Y.}~\bibnamefont{Chen}},
  \bibinfo{author}{\bibfnamefont{B.}~\bibnamefont{Fornal}},
  \bibinfo{author}{\bibfnamefont{P.}~\bibnamefont{Sandick}},
  \bibinfo{author}{\bibfnamefont{J.}~\bibnamefont{Shu}},
  \bibinfo{author}{\bibfnamefont{X.}~\bibnamefont{Xue}},
  \bibinfo{author}{\bibfnamefont{Y.}~\bibnamefont{Zhao}}, \bibnamefont{and}
  \bibinfo{author}{\bibfnamefont{J.}~\bibnamefont{Zong}}
  (\bibinfo{year}{2021}), \eprint{2110.09685}.

\bibitem[{\citenamefont{Hochberg
  et~al.}(2016{\natexlab{a}})\citenamefont{Hochberg, Zhao, and
  Zurek}}]{Hochberg:2015pha}
\bibinfo{author}{\bibfnamefont{Y.}~\bibnamefont{Hochberg}},
  \bibinfo{author}{\bibfnamefont{Y.}~\bibnamefont{Zhao}}, \bibnamefont{and}
  \bibinfo{author}{\bibfnamefont{K.~M.} \bibnamefont{Zurek}},
  \bibinfo{journal}{Phys. Rev. Lett.} \textbf{\bibinfo{volume}{116}},
  \bibinfo{pages}{011301} (\bibinfo{year}{2016}{\natexlab{a}}),
  \eprint{1504.07237}.

\bibitem[{\citenamefont{Hochberg
  et~al.}(2016{\natexlab{b}})\citenamefont{Hochberg, Pyle, Zhao, and
  Zurek}}]{Hochberg:2015fth}
\bibinfo{author}{\bibfnamefont{Y.}~\bibnamefont{Hochberg}},
  \bibinfo{author}{\bibfnamefont{M.}~\bibnamefont{Pyle}},
  \bibinfo{author}{\bibfnamefont{Y.}~\bibnamefont{Zhao}}, \bibnamefont{and}
  \bibinfo{author}{\bibfnamefont{K.~M.} \bibnamefont{Zurek}},
  \bibinfo{journal}{JHEP} \textbf{\bibinfo{volume}{08}}, \bibinfo{pages}{057}
  (\bibinfo{year}{2016}{\natexlab{b}}), \eprint{1512.04533}.

\bibitem[{\citenamefont{Hochberg et~al.}(2019)\citenamefont{Hochberg, Charaev,
  Nam, Verma, Colangelo, and Berggren}}]{Hochberg:2019cyy}
\bibinfo{author}{\bibfnamefont{Y.}~\bibnamefont{Hochberg}},
  \bibinfo{author}{\bibfnamefont{I.}~\bibnamefont{Charaev}},
  \bibinfo{author}{\bibfnamefont{S.-W.} \bibnamefont{Nam}},
  \bibinfo{author}{\bibfnamefont{V.}~\bibnamefont{Verma}},
  \bibinfo{author}{\bibfnamefont{M.}~\bibnamefont{Colangelo}},
  \bibnamefont{and} \bibinfo{author}{\bibfnamefont{K.~K.}
  \bibnamefont{Berggren}}, \bibinfo{journal}{Phys. Rev. Lett.}
  \textbf{\bibinfo{volume}{123}}, \bibinfo{pages}{151802}
  (\bibinfo{year}{2019}), \eprint{1903.05101}.

\bibitem[{\citenamefont{Graham et~al.}(2012)\citenamefont{Graham, Kaplan,
  Rajendran, and Walters}}]{Graham:2012su}
\bibinfo{author}{\bibfnamefont{P.~W.} \bibnamefont{Graham}},
  \bibinfo{author}{\bibfnamefont{D.~E.} \bibnamefont{Kaplan}},
  \bibinfo{author}{\bibfnamefont{S.}~\bibnamefont{Rajendran}},
  \bibnamefont{and} \bibinfo{author}{\bibfnamefont{M.~T.}
  \bibnamefont{Walters}}, \bibinfo{journal}{Phys. Dark Univ.}
  \textbf{\bibinfo{volume}{1}}, \bibinfo{pages}{32} (\bibinfo{year}{2012}),
  \eprint{1203.2531}.

\bibitem[{\citenamefont{Lee et~al.}(2015)\citenamefont{Lee, Lisanti,
  Mishra-Sharma, and Safdi}}]{Lee:2015qva}
\bibinfo{author}{\bibfnamefont{S.~K.} \bibnamefont{Lee}},
  \bibinfo{author}{\bibfnamefont{M.}~\bibnamefont{Lisanti}},
  \bibinfo{author}{\bibfnamefont{S.}~\bibnamefont{Mishra-Sharma}},
  \bibnamefont{and} \bibinfo{author}{\bibfnamefont{B.~R.} \bibnamefont{Safdi}},
  \bibinfo{journal}{Phys. Rev. D} \textbf{\bibinfo{volume}{92}},
  \bibinfo{pages}{083517} (\bibinfo{year}{2015}), \eprint{1508.07361}.

\bibitem[{\citenamefont{Essig et~al.}(2016)\citenamefont{Essig,
  Fernandez-Serra, Mardon, Soto, Volansky, and Yu}}]{Essig:2015cda}
\bibinfo{author}{\bibfnamefont{R.}~\bibnamefont{Essig}},
  \bibinfo{author}{\bibfnamefont{M.}~\bibnamefont{Fernandez-Serra}},
  \bibinfo{author}{\bibfnamefont{J.}~\bibnamefont{Mardon}},
  \bibinfo{author}{\bibfnamefont{A.}~\bibnamefont{Soto}},
  \bibinfo{author}{\bibfnamefont{T.}~\bibnamefont{Volansky}}, \bibnamefont{and}
  \bibinfo{author}{\bibfnamefont{T.-T.} \bibnamefont{Yu}},
  \bibinfo{journal}{JHEP} \textbf{\bibinfo{volume}{05}}, \bibinfo{pages}{046}
  (\bibinfo{year}{2016}), \eprint{1509.01598}.

\bibitem[{\citenamefont{Crisler et~al.}(2018)\citenamefont{Crisler, Essig,
  Estrada, Fernandez, Tiffenberg, Sofo~haro, Volansky, and
  Yu}}]{Crisler:2018gci}
\bibinfo{author}{\bibfnamefont{M.}~\bibnamefont{Crisler}},
  \bibinfo{author}{\bibfnamefont{R.}~\bibnamefont{Essig}},
  \bibinfo{author}{\bibfnamefont{J.}~\bibnamefont{Estrada}},
  \bibinfo{author}{\bibfnamefont{G.}~\bibnamefont{Fernandez}},
  \bibinfo{author}{\bibfnamefont{J.}~\bibnamefont{Tiffenberg}},
  \bibinfo{author}{\bibfnamefont{M.}~\bibnamefont{Sofo~haro}},
  \bibinfo{author}{\bibfnamefont{T.}~\bibnamefont{Volansky}}, \bibnamefont{and}
  \bibinfo{author}{\bibfnamefont{T.-T.} \bibnamefont{Yu}}
  (\bibinfo{collaboration}{SENSEI}), \bibinfo{journal}{Phys. Rev. Lett.}
  \textbf{\bibinfo{volume}{121}}, \bibinfo{pages}{061803}
  (\bibinfo{year}{2018}), \eprint{1804.00088}.

\bibitem[{\citenamefont{Agnese et~al.}(2018)}]{SuperCDMS:2018mne}
\bibinfo{author}{\bibfnamefont{R.}~\bibnamefont{Agnese}} \bibnamefont{et~al.}
  (\bibinfo{collaboration}{SuperCDMS}), \bibinfo{journal}{Phys. Rev. Lett.}
  \textbf{\bibinfo{volume}{121}}, \bibinfo{pages}{051301}
  (\bibinfo{year}{2018}), \bibinfo{note}{[Erratum: Phys.Rev.Lett. 122, 069901
  (2019)]}, \eprint{1804.10697}.

\bibitem[{\citenamefont{Abramoff et~al.}(2019)}]{SENSEI:2019ibb}
\bibinfo{author}{\bibfnamefont{O.}~\bibnamefont{Abramoff}} \bibnamefont{et~al.}
  (\bibinfo{collaboration}{SENSEI}), \bibinfo{journal}{Phys. Rev. Lett.}
  \textbf{\bibinfo{volume}{122}}, \bibinfo{pages}{161801}
  (\bibinfo{year}{2019}), \eprint{1901.10478}.

\bibitem[{\citenamefont{Liu et~al.}(2019)}]{CDEX:2019hzn}
\bibinfo{author}{\bibfnamefont{Z.~Z.} \bibnamefont{Liu}} \bibnamefont{et~al.}
  (\bibinfo{collaboration}{CDEX}), \bibinfo{journal}{Phys. Rev. Lett.}
  \textbf{\bibinfo{volume}{123}}, \bibinfo{pages}{161301}
  (\bibinfo{year}{2019}), \eprint{1905.00354}.

\bibitem[{\citenamefont{Aguilar-Arevalo et~al.}(2019)}]{DAMIC:2019dcn}
\bibinfo{author}{\bibfnamefont{A.}~\bibnamefont{Aguilar-Arevalo}}
  \bibnamefont{et~al.} (\bibinfo{collaboration}{DAMIC}),
  \bibinfo{journal}{Phys. Rev. Lett.} \textbf{\bibinfo{volume}{123}},
  \bibinfo{pages}{181802} (\bibinfo{year}{2019}), \eprint{1907.12628}.

\bibitem[{\citenamefont{Andersson et~al.}(2020)\citenamefont{Andersson,
  B\"okmark, Catena, Emken, Moberg, and \r{A}strand}}]{Andersson:2020uwc}
\bibinfo{author}{\bibfnamefont{E.}~\bibnamefont{Andersson}},
  \bibinfo{author}{\bibfnamefont{A.}~\bibnamefont{B\"okmark}},
  \bibinfo{author}{\bibfnamefont{R.}~\bibnamefont{Catena}},
  \bibinfo{author}{\bibfnamefont{T.}~\bibnamefont{Emken}},
  \bibinfo{author}{\bibfnamefont{H.~K.} \bibnamefont{Moberg}},
  \bibnamefont{and}
  \bibinfo{author}{\bibfnamefont{E.}~\bibnamefont{\r{A}strand}},
  \bibinfo{journal}{JCAP} \textbf{\bibinfo{volume}{05}}, \bibinfo{pages}{036}
  (\bibinfo{year}{2020}), \eprint{2001.08910}.

\bibitem[{\citenamefont{Barak et~al.}(2020)}]{SENSEI:2020dpa}
\bibinfo{author}{\bibfnamefont{L.}~\bibnamefont{Barak}} \bibnamefont{et~al.}
  (\bibinfo{collaboration}{SENSEI}), \bibinfo{journal}{Phys. Rev. Lett.}
  \textbf{\bibinfo{volume}{125}}, \bibinfo{pages}{171802}
  (\bibinfo{year}{2020}), \eprint{2004.11378}.

\bibitem[{\citenamefont{Amaral et~al.}(2020)}]{SuperCDMS:2020ymb}
\bibinfo{author}{\bibfnamefont{D.~W.} \bibnamefont{Amaral}}
  \bibnamefont{et~al.} (\bibinfo{collaboration}{SuperCDMS}),
  \bibinfo{journal}{Phys. Rev. D} \textbf{\bibinfo{volume}{102}},
  \bibinfo{pages}{091101} (\bibinfo{year}{2020}), \eprint{2005.14067}.

\bibitem[{\citenamefont{Catena et~al.}(2021)\citenamefont{Catena, Emken, Matas,
  Spaldin, and Urdshals}}]{Catena:2021qsr}
\bibinfo{author}{\bibfnamefont{R.}~\bibnamefont{Catena}},
  \bibinfo{author}{\bibfnamefont{T.}~\bibnamefont{Emken}},
  \bibinfo{author}{\bibfnamefont{M.}~\bibnamefont{Matas}},
  \bibinfo{author}{\bibfnamefont{N.~A.} \bibnamefont{Spaldin}},
  \bibnamefont{and} \bibinfo{author}{\bibfnamefont{E.}~\bibnamefont{Urdshals}},
  \bibinfo{journal}{Phys. Rev. Res.} \textbf{\bibinfo{volume}{3}},
  \bibinfo{pages}{033149} (\bibinfo{year}{2021}), \eprint{2105.02233}.

\bibitem[{\citenamefont{Agnes et~al.}(2018)}]{DarkSide:2018ppu}
\bibinfo{author}{\bibfnamefont{P.}~\bibnamefont{Agnes}} \bibnamefont{et~al.}
  (\bibinfo{collaboration}{DarkSide}), \bibinfo{journal}{Phys. Rev. Lett.}
  \textbf{\bibinfo{volume}{121}}, \bibinfo{pages}{111303}
  (\bibinfo{year}{2018}), \eprint{1802.06998}.

\bibitem[{\citenamefont{Essig et~al.}(2012{\natexlab{b}})\citenamefont{Essig,
  Manalaysay, Mardon, Sorensen, and Volansky}}]{Essig:2012yx}
\bibinfo{author}{\bibfnamefont{R.}~\bibnamefont{Essig}},
  \bibinfo{author}{\bibfnamefont{A.}~\bibnamefont{Manalaysay}},
  \bibinfo{author}{\bibfnamefont{J.}~\bibnamefont{Mardon}},
  \bibinfo{author}{\bibfnamefont{P.}~\bibnamefont{Sorensen}}, \bibnamefont{and}
  \bibinfo{author}{\bibfnamefont{T.}~\bibnamefont{Volansky}},
  \bibinfo{journal}{Phys. Rev. Lett.} \textbf{\bibinfo{volume}{109}},
  \bibinfo{pages}{021301} (\bibinfo{year}{2012}{\natexlab{b}}),
  \eprint{1206.2644}.

\bibitem[{\citenamefont{Essig et~al.}(2017)\citenamefont{Essig, Volansky, and
  Yu}}]{Essig:2017kqs}
\bibinfo{author}{\bibfnamefont{R.}~\bibnamefont{Essig}},
  \bibinfo{author}{\bibfnamefont{T.}~\bibnamefont{Volansky}}, \bibnamefont{and}
  \bibinfo{author}{\bibfnamefont{T.-T.} \bibnamefont{Yu}},
  \bibinfo{journal}{Phys. Rev. D} \textbf{\bibinfo{volume}{96}},
  \bibinfo{pages}{043017} (\bibinfo{year}{2017}), \eprint{1703.00910}.

\bibitem[{\citenamefont{Cui et~al.}(2017)}]{PandaX-II:2017hlx}
\bibinfo{author}{\bibfnamefont{X.}~\bibnamefont{Cui}} \bibnamefont{et~al.}
  (\bibinfo{collaboration}{PandaX-II}), \bibinfo{journal}{Phys. Rev. Lett.}
  \textbf{\bibinfo{volume}{119}}, \bibinfo{pages}{181302}
  (\bibinfo{year}{2017}), \eprint{1708.06917}.

\bibitem[{\citenamefont{Aprile et~al.}(2019)}]{XENON:2019gfn}
\bibinfo{author}{\bibfnamefont{E.}~\bibnamefont{Aprile}} \bibnamefont{et~al.}
  (\bibinfo{collaboration}{XENON}), \bibinfo{journal}{Phys. Rev. Lett.}
  \textbf{\bibinfo{volume}{123}}, \bibinfo{pages}{251801}
  (\bibinfo{year}{2019}), \eprint{1907.11485}.

\bibitem[{\citenamefont{Derenzo et~al.}(2017)\citenamefont{Derenzo, Essig,
  Massari, Soto, and Yu}}]{Derenzo:2016fse}
\bibinfo{author}{\bibfnamefont{S.}~\bibnamefont{Derenzo}},
  \bibinfo{author}{\bibfnamefont{R.}~\bibnamefont{Essig}},
  \bibinfo{author}{\bibfnamefont{A.}~\bibnamefont{Massari}},
  \bibinfo{author}{\bibfnamefont{A.}~\bibnamefont{Soto}}, \bibnamefont{and}
  \bibinfo{author}{\bibfnamefont{T.-T.} \bibnamefont{Yu}},
  \bibinfo{journal}{Phys. Rev. D} \textbf{\bibinfo{volume}{96}},
  \bibinfo{pages}{016026} (\bibinfo{year}{2017}), \eprint{1607.01009}.

\bibitem[{\citenamefont{Blanco et~al.}(2020)\citenamefont{Blanco, Collar, Kahn,
  and Lillard}}]{Blanco:2019lrf}
\bibinfo{author}{\bibfnamefont{C.}~\bibnamefont{Blanco}},
  \bibinfo{author}{\bibfnamefont{J.~I.} \bibnamefont{Collar}},
  \bibinfo{author}{\bibfnamefont{Y.}~\bibnamefont{Kahn}}, \bibnamefont{and}
  \bibinfo{author}{\bibfnamefont{B.}~\bibnamefont{Lillard}},
  \bibinfo{journal}{Phys. Rev. D} \textbf{\bibinfo{volume}{101}},
  \bibinfo{pages}{056001} (\bibinfo{year}{2020}), \eprint{1912.02822}.

\bibitem[{\citenamefont{Hochberg et~al.}(2017)\citenamefont{Hochberg, Kahn,
  Lisanti, Tully, and Zurek}}]{Hochberg:2016ntt}
\bibinfo{author}{\bibfnamefont{Y.}~\bibnamefont{Hochberg}},
  \bibinfo{author}{\bibfnamefont{Y.}~\bibnamefont{Kahn}},
  \bibinfo{author}{\bibfnamefont{M.}~\bibnamefont{Lisanti}},
  \bibinfo{author}{\bibfnamefont{C.~G.} \bibnamefont{Tully}}, \bibnamefont{and}
  \bibinfo{author}{\bibfnamefont{K.~M.} \bibnamefont{Zurek}},
  \bibinfo{journal}{Phys. Lett. B} \textbf{\bibinfo{volume}{772}},
  \bibinfo{pages}{239} (\bibinfo{year}{2017}), \eprint{1606.08849}.

\bibitem[{\citenamefont{Geilhufe et~al.}(2018)\citenamefont{Geilhufe,
  Olsthoorn, Ferella, Koski, Kahlhoefer, Conrad, and
  Balatsky}}]{Geilhufe:2018gry}
\bibinfo{author}{\bibfnamefont{R.~M.} \bibnamefont{Geilhufe}},
  \bibinfo{author}{\bibfnamefont{B.}~\bibnamefont{Olsthoorn}},
  \bibinfo{author}{\bibfnamefont{A.}~\bibnamefont{Ferella}},
  \bibinfo{author}{\bibfnamefont{T.}~\bibnamefont{Koski}},
  \bibinfo{author}{\bibfnamefont{F.}~\bibnamefont{Kahlhoefer}},
  \bibinfo{author}{\bibfnamefont{J.}~\bibnamefont{Conrad}}, \bibnamefont{and}
  \bibinfo{author}{\bibfnamefont{A.~V.} \bibnamefont{Balatsky}},
  \bibinfo{journal}{Phys. Status Solidi RRL} \textbf{\bibinfo{volume}{12}},
  \bibinfo{pages}{1800293} (\bibinfo{year}{2018}), \eprint{1806.06040}.

\bibitem[{\citenamefont{Hochberg et~al.}(2018)\citenamefont{Hochberg, Kahn,
  Lisanti, Zurek, Grushin, Ilan, Griffin, Liu, Weber, and
  Neaton}}]{Hochberg:2017wce}
\bibinfo{author}{\bibfnamefont{Y.}~\bibnamefont{Hochberg}},
  \bibinfo{author}{\bibfnamefont{Y.}~\bibnamefont{Kahn}},
  \bibinfo{author}{\bibfnamefont{M.}~\bibnamefont{Lisanti}},
  \bibinfo{author}{\bibfnamefont{K.~M.} \bibnamefont{Zurek}},
  \bibinfo{author}{\bibfnamefont{A.~G.} \bibnamefont{Grushin}},
  \bibinfo{author}{\bibfnamefont{R.}~\bibnamefont{Ilan}},
  \bibinfo{author}{\bibfnamefont{S.~M.} \bibnamefont{Griffin}},
  \bibinfo{author}{\bibfnamefont{Z.-F.} \bibnamefont{Liu}},
  \bibinfo{author}{\bibfnamefont{S.~F.} \bibnamefont{Weber}}, \bibnamefont{and}
  \bibinfo{author}{\bibfnamefont{J.~B.} \bibnamefont{Neaton}},
  \bibinfo{journal}{Phys. Rev. D} \textbf{\bibinfo{volume}{97}},
  \bibinfo{pages}{015004} (\bibinfo{year}{2018}), \eprint{1708.08929}.

\bibitem[{\citenamefont{Geilhufe et~al.}(2020)\citenamefont{Geilhufe,
  Kahlhoefer, and Winkler}}]{Geilhufe:2019ndy}
\bibinfo{author}{\bibfnamefont{R.~M.} \bibnamefont{Geilhufe}},
  \bibinfo{author}{\bibfnamefont{F.}~\bibnamefont{Kahlhoefer}},
  \bibnamefont{and} \bibinfo{author}{\bibfnamefont{M.~W.}
  \bibnamefont{Winkler}}, \bibinfo{journal}{Phys. Rev. D}
  \textbf{\bibinfo{volume}{101}}, \bibinfo{pages}{055005}
  (\bibinfo{year}{2020}), \eprint{1910.02091}.

\bibitem[{\citenamefont{Knapen et~al.}(2018)\citenamefont{Knapen, Lin, Pyle,
  and Zurek}}]{Knapen:2017ekk}
\bibinfo{author}{\bibfnamefont{S.}~\bibnamefont{Knapen}},
  \bibinfo{author}{\bibfnamefont{T.}~\bibnamefont{Lin}},
  \bibinfo{author}{\bibfnamefont{M.}~\bibnamefont{Pyle}}, \bibnamefont{and}
  \bibinfo{author}{\bibfnamefont{K.~M.} \bibnamefont{Zurek}},
  \bibinfo{journal}{Phys. Lett. B} \textbf{\bibinfo{volume}{785}},
  \bibinfo{pages}{386} (\bibinfo{year}{2018}), \eprint{1712.06598}.

\bibitem[{\citenamefont{Boehm et~al.}(2004)\citenamefont{Boehm, Fayet, and
  Silk}}]{Boehm:2003ha}
\bibinfo{author}{\bibfnamefont{C.}~\bibnamefont{Boehm}},
  \bibinfo{author}{\bibfnamefont{P.}~\bibnamefont{Fayet}}, \bibnamefont{and}
  \bibinfo{author}{\bibfnamefont{J.}~\bibnamefont{Silk}},
  \bibinfo{journal}{Phys. Rev. D} \textbf{\bibinfo{volume}{69}},
  \bibinfo{pages}{101302} (\bibinfo{year}{2004}), \eprint{hep-ph/0311143}.

\bibitem[{\citenamefont{Borodatchenkova
  et~al.}(2006)\citenamefont{Borodatchenkova, Choudhury, and
  Drees}}]{Borodatchenkova:2005ct}
\bibinfo{author}{\bibfnamefont{N.}~\bibnamefont{Borodatchenkova}},
  \bibinfo{author}{\bibfnamefont{D.}~\bibnamefont{Choudhury}},
  \bibnamefont{and} \bibinfo{author}{\bibfnamefont{M.}~\bibnamefont{Drees}},
  \bibinfo{journal}{Phys. Rev. Lett.} \textbf{\bibinfo{volume}{96}},
  \bibinfo{pages}{141802} (\bibinfo{year}{2006}), \eprint{hep-ph/0510147}.

\bibitem[{\citenamefont{Pospelov
  et~al.}(2008{\natexlab{a}})\citenamefont{Pospelov, Ritz, and
  Voloshin}}]{Pospelov:2007mp}
\bibinfo{author}{\bibfnamefont{M.}~\bibnamefont{Pospelov}},
  \bibinfo{author}{\bibfnamefont{A.}~\bibnamefont{Ritz}}, \bibnamefont{and}
  \bibinfo{author}{\bibfnamefont{M.~B.} \bibnamefont{Voloshin}},
  \bibinfo{journal}{Phys. Lett. B} \textbf{\bibinfo{volume}{662}},
  \bibinfo{pages}{53} (\bibinfo{year}{2008}{\natexlab{a}}), \eprint{0711.4866}.

\bibitem[{\citenamefont{Hooper and Zurek}(2008)}]{Hooper:2008im}
\bibinfo{author}{\bibfnamefont{D.}~\bibnamefont{Hooper}} \bibnamefont{and}
  \bibinfo{author}{\bibfnamefont{K.~M.} \bibnamefont{Zurek}},
  \bibinfo{journal}{Phys. Rev. D} \textbf{\bibinfo{volume}{77}},
  \bibinfo{pages}{087302} (\bibinfo{year}{2008}), \eprint{0801.3686}.

\bibitem[{\citenamefont{Feng and Kumar}(2008)}]{Feng:2008ya}
\bibinfo{author}{\bibfnamefont{J.~L.} \bibnamefont{Feng}} \bibnamefont{and}
  \bibinfo{author}{\bibfnamefont{J.}~\bibnamefont{Kumar}},
  \bibinfo{journal}{Phys. Rev. Lett.} \textbf{\bibinfo{volume}{101}},
  \bibinfo{pages}{231301} (\bibinfo{year}{2008}), \eprint{0803.4196}.

\bibitem[{\citenamefont{Pospelov
  et~al.}(2008{\natexlab{b}})\citenamefont{Pospelov, Ritz, and
  Voloshin}}]{Pospelov:2008jk}
\bibinfo{author}{\bibfnamefont{M.}~\bibnamefont{Pospelov}},
  \bibinfo{author}{\bibfnamefont{A.}~\bibnamefont{Ritz}}, \bibnamefont{and}
  \bibinfo{author}{\bibfnamefont{M.~B.} \bibnamefont{Voloshin}},
  \bibinfo{journal}{Phys. Rev. D} \textbf{\bibinfo{volume}{78}},
  \bibinfo{pages}{115012} (\bibinfo{year}{2008}{\natexlab{b}}),
  \eprint{0807.3279}.

\bibitem[{\citenamefont{Falkowski et~al.}(2011)\citenamefont{Falkowski,
  Ruderman, and Volansky}}]{Falkowski:2011xh}
\bibinfo{author}{\bibfnamefont{A.}~\bibnamefont{Falkowski}},
  \bibinfo{author}{\bibfnamefont{J.~T.} \bibnamefont{Ruderman}},
  \bibnamefont{and} \bibinfo{author}{\bibfnamefont{T.}~\bibnamefont{Volansky}},
  \bibinfo{journal}{JHEP} \textbf{\bibinfo{volume}{05}}, \bibinfo{pages}{106}
  (\bibinfo{year}{2011}), \eprint{1101.4936}.

\bibitem[{\citenamefont{Tucker-Smith and Weiner}(2001)}]{Tucker-Smith:2001myb}
\bibinfo{author}{\bibfnamefont{D.}~\bibnamefont{Tucker-Smith}}
  \bibnamefont{and} \bibinfo{author}{\bibfnamefont{N.}~\bibnamefont{Weiner}},
  \bibinfo{journal}{Phys. Rev. D} \textbf{\bibinfo{volume}{64}},
  \bibinfo{pages}{043502} (\bibinfo{year}{2001}), \eprint{hep-ph/0101138}.

\bibitem[{\citenamefont{Tucker-Smith and Weiner}(2005)}]{Tucker-Smith:2004mxa}
\bibinfo{author}{\bibfnamefont{D.}~\bibnamefont{Tucker-Smith}}
  \bibnamefont{and} \bibinfo{author}{\bibfnamefont{N.}~\bibnamefont{Weiner}},
  \bibinfo{journal}{Phys. Rev. D} \textbf{\bibinfo{volume}{72}},
  \bibinfo{pages}{063509} (\bibinfo{year}{2005}), \eprint{hep-ph/0402065}.

\bibitem[{\citenamefont{Finkbeiner and Weiner}(2007)}]{Finkbeiner:2007kk}
\bibinfo{author}{\bibfnamefont{D.~P.} \bibnamefont{Finkbeiner}}
  \bibnamefont{and} \bibinfo{author}{\bibfnamefont{N.}~\bibnamefont{Weiner}},
  \bibinfo{journal}{Phys. Rev. D} \textbf{\bibinfo{volume}{76}},
  \bibinfo{pages}{083519} (\bibinfo{year}{2007}), \eprint{astro-ph/0702587}.

\bibitem[{\citenamefont{Arina and Fornengo}(2007)}]{Arina:2007tm}
\bibinfo{author}{\bibfnamefont{C.}~\bibnamefont{Arina}} \bibnamefont{and}
  \bibinfo{author}{\bibfnamefont{N.}~\bibnamefont{Fornengo}},
  \bibinfo{journal}{JHEP} \textbf{\bibinfo{volume}{11}}, \bibinfo{pages}{029}
  (\bibinfo{year}{2007}), \eprint{0709.4477}.

\bibitem[{\citenamefont{Chang et~al.}(2009)\citenamefont{Chang, Kribs,
  Tucker-Smith, and Weiner}}]{Chang:2008gd}
\bibinfo{author}{\bibfnamefont{S.}~\bibnamefont{Chang}},
  \bibinfo{author}{\bibfnamefont{G.~D.} \bibnamefont{Kribs}},
  \bibinfo{author}{\bibfnamefont{D.}~\bibnamefont{Tucker-Smith}},
  \bibnamefont{and} \bibinfo{author}{\bibfnamefont{N.}~\bibnamefont{Weiner}},
  \bibinfo{journal}{Phys. Rev. D} \textbf{\bibinfo{volume}{79}},
  \bibinfo{pages}{043513} (\bibinfo{year}{2009}), \eprint{0807.2250}.

\bibitem[{\citenamefont{Cui et~al.}(2009)\citenamefont{Cui, Morrissey, Poland,
  and Randall}}]{Cui:2009xq}
\bibinfo{author}{\bibfnamefont{Y.}~\bibnamefont{Cui}},
  \bibinfo{author}{\bibfnamefont{D.~E.} \bibnamefont{Morrissey}},
  \bibinfo{author}{\bibfnamefont{D.}~\bibnamefont{Poland}}, \bibnamefont{and}
  \bibinfo{author}{\bibfnamefont{L.}~\bibnamefont{Randall}},
  \bibinfo{journal}{JHEP} \textbf{\bibinfo{volume}{05}}, \bibinfo{pages}{076}
  (\bibinfo{year}{2009}), \eprint{0901.0557}.

\bibitem[{\citenamefont{Lin and Finkbeiner}(2011)}]{Lin:2010sb}
\bibinfo{author}{\bibfnamefont{T.}~\bibnamefont{Lin}} \bibnamefont{and}
  \bibinfo{author}{\bibfnamefont{D.~P.} \bibnamefont{Finkbeiner}},
  \bibinfo{journal}{Phys. Rev. D} \textbf{\bibinfo{volume}{83}},
  \bibinfo{pages}{083510} (\bibinfo{year}{2011}), \eprint{1011.3052}.

\bibitem[{\citenamefont{De~Simone et~al.}(2010)\citenamefont{De~Simone, Sanz,
  and Sato}}]{DeSimone:2010tf}
\bibinfo{author}{\bibfnamefont{A.}~\bibnamefont{De~Simone}},
  \bibinfo{author}{\bibfnamefont{V.}~\bibnamefont{Sanz}}, \bibnamefont{and}
  \bibinfo{author}{\bibfnamefont{H.~P.} \bibnamefont{Sato}},
  \bibinfo{journal}{Phys. Rev. Lett.} \textbf{\bibinfo{volume}{105}},
  \bibinfo{pages}{121802} (\bibinfo{year}{2010}), \eprint{1004.1567}.

\bibitem[{\citenamefont{An et~al.}(2012)\citenamefont{An, Dev, Cai, and
  Mohapatra}}]{An:2011uq}
\bibinfo{author}{\bibfnamefont{H.}~\bibnamefont{An}},
  \bibinfo{author}{\bibfnamefont{P.~S.~B.} \bibnamefont{Dev}},
  \bibinfo{author}{\bibfnamefont{Y.}~\bibnamefont{Cai}}, \bibnamefont{and}
  \bibinfo{author}{\bibfnamefont{R.~N.} \bibnamefont{Mohapatra}},
  \bibinfo{journal}{Phys. Rev. Lett.} \textbf{\bibinfo{volume}{108}},
  \bibinfo{pages}{081806} (\bibinfo{year}{2012}), \eprint{1110.1366}.

\bibitem[{\citenamefont{Pospelov et~al.}(2014)\citenamefont{Pospelov, Weiner,
  and Yavin}}]{Pospelov:2013nea}
\bibinfo{author}{\bibfnamefont{M.}~\bibnamefont{Pospelov}},
  \bibinfo{author}{\bibfnamefont{N.}~\bibnamefont{Weiner}}, \bibnamefont{and}
  \bibinfo{author}{\bibfnamefont{I.}~\bibnamefont{Yavin}},
  \bibinfo{journal}{Phys. Rev. D} \textbf{\bibinfo{volume}{89}},
  \bibinfo{pages}{055008} (\bibinfo{year}{2014}), \eprint{1312.1363}.

\bibitem[{\citenamefont{Finkbeiner and Weiner}(2016)}]{Finkbeiner:2014sja}
\bibinfo{author}{\bibfnamefont{D.~P.} \bibnamefont{Finkbeiner}}
  \bibnamefont{and} \bibinfo{author}{\bibfnamefont{N.}~\bibnamefont{Weiner}},
  \bibinfo{journal}{Phys. Rev. D} \textbf{\bibinfo{volume}{94}},
  \bibinfo{pages}{083002} (\bibinfo{year}{2016}), \eprint{1402.6671}.

\bibitem[{\citenamefont{Dienes et~al.}(2015)\citenamefont{Dienes, Kumar,
  Thomas, and Yaylali}}]{Dienes:2014via}
\bibinfo{author}{\bibfnamefont{K.~R.} \bibnamefont{Dienes}},
  \bibinfo{author}{\bibfnamefont{J.}~\bibnamefont{Kumar}},
  \bibinfo{author}{\bibfnamefont{B.}~\bibnamefont{Thomas}}, \bibnamefont{and}
  \bibinfo{author}{\bibfnamefont{D.}~\bibnamefont{Yaylali}},
  \bibinfo{journal}{Phys. Rev. Lett.} \textbf{\bibinfo{volume}{114}},
  \bibinfo{pages}{051301} (\bibinfo{year}{2015}), \eprint{1406.4868}.

\bibitem[{\citenamefont{Dror et~al.}(2020{\natexlab{a}})\citenamefont{Dror,
  Elor, and Mcgehee}}]{Dror:2019onn}
\bibinfo{author}{\bibfnamefont{J.~A.} \bibnamefont{Dror}},
  \bibinfo{author}{\bibfnamefont{G.}~\bibnamefont{Elor}}, \bibnamefont{and}
  \bibinfo{author}{\bibfnamefont{R.}~\bibnamefont{Mcgehee}},
  \bibinfo{journal}{Phys. Rev. Lett.} \textbf{\bibinfo{volume}{124}},
  \bibinfo{pages}{18} (\bibinfo{year}{2020}{\natexlab{a}}),
  \eprint{1905.12635}.

\bibitem[{\citenamefont{Dror et~al.}(2020{\natexlab{b}})\citenamefont{Dror,
  Elor, and Mcgehee}}]{Dror:2019dib}
\bibinfo{author}{\bibfnamefont{J.~A.} \bibnamefont{Dror}},
  \bibinfo{author}{\bibfnamefont{G.}~\bibnamefont{Elor}}, \bibnamefont{and}
  \bibinfo{author}{\bibfnamefont{R.}~\bibnamefont{Mcgehee}},
  \bibinfo{journal}{JHEP} \textbf{\bibinfo{volume}{02}}, \bibinfo{pages}{134}
  (\bibinfo{year}{2020}{\natexlab{b}}), \eprint{1908.10861}.

\bibitem[{\citenamefont{Carrillo~Gonz\'alez and
  Toro}(2021)}]{CarrilloGonzalez:2021lxm}
\bibinfo{author}{\bibfnamefont{M.}~\bibnamefont{Carrillo~Gonz\'alez}}
  \bibnamefont{and} \bibinfo{author}{\bibfnamefont{N.}~\bibnamefont{Toro}}
  (\bibinfo{year}{2021}), \eprint{2108.13422}.

\bibitem[{\citenamefont{Guo et~al.}(2021)\citenamefont{Guo, He, Liu, and
  Wang}}]{Guo:2021vpb}
\bibinfo{author}{\bibfnamefont{J.}~\bibnamefont{Guo}},
  \bibinfo{author}{\bibfnamefont{Y.}~\bibnamefont{He}},
  \bibinfo{author}{\bibfnamefont{J.}~\bibnamefont{Liu}}, \bibnamefont{and}
  \bibinfo{author}{\bibfnamefont{X.-P.} \bibnamefont{Wang}}
  (\bibinfo{year}{2021}), \eprint{2111.01164}.

\bibitem[{\citenamefont{Harigaya et~al.}(2020)\citenamefont{Harigaya, Nakai,
  and Suzuki}}]{Harigaya:2020ckz}
\bibinfo{author}{\bibfnamefont{K.}~\bibnamefont{Harigaya}},
  \bibinfo{author}{\bibfnamefont{Y.}~\bibnamefont{Nakai}}, \bibnamefont{and}
  \bibinfo{author}{\bibfnamefont{M.}~\bibnamefont{Suzuki}},
  \bibinfo{journal}{Phys. Lett. B} \textbf{\bibinfo{volume}{809}},
  \bibinfo{pages}{135729} (\bibinfo{year}{2020}), \eprint{2006.11938}.

\bibitem[{\citenamefont{Lee}(2021)}]{Lee:2020wmh}
\bibinfo{author}{\bibfnamefont{H.~M.} \bibnamefont{Lee}},
  \bibinfo{journal}{JHEP} \textbf{\bibinfo{volume}{01}}, \bibinfo{pages}{019}
  (\bibinfo{year}{2021}), \eprint{2006.13183}.

\bibitem[{\citenamefont{Baryakhtar et~al.}(2020)\citenamefont{Baryakhtar,
  Berlin, Liu, and Weiner}}]{Baryakhtar:2020rwy}
\bibinfo{author}{\bibfnamefont{M.}~\bibnamefont{Baryakhtar}},
  \bibinfo{author}{\bibfnamefont{A.}~\bibnamefont{Berlin}},
  \bibinfo{author}{\bibfnamefont{H.}~\bibnamefont{Liu}}, \bibnamefont{and}
  \bibinfo{author}{\bibfnamefont{N.}~\bibnamefont{Weiner}}
  (\bibinfo{year}{2020}), \eprint{2006.13918}.

\bibitem[{\citenamefont{Bramante and Song}(2020)}]{Bramante:2020zos}
\bibinfo{author}{\bibfnamefont{J.}~\bibnamefont{Bramante}} \bibnamefont{and}
  \bibinfo{author}{\bibfnamefont{N.}~\bibnamefont{Song}},
  \bibinfo{journal}{Phys. Rev. Lett.} \textbf{\bibinfo{volume}{125}},
  \bibinfo{pages}{161805} (\bibinfo{year}{2020}), \eprint{2006.14089}.

\bibitem[{\citenamefont{Choi et~al.}(2021)\citenamefont{Choi, Lee, and
  Zhu}}]{Choi:2020ysq}
\bibinfo{author}{\bibfnamefont{S.-M.} \bibnamefont{Choi}},
  \bibinfo{author}{\bibfnamefont{H.~M.} \bibnamefont{Lee}}, \bibnamefont{and}
  \bibinfo{author}{\bibfnamefont{B.}~\bibnamefont{Zhu}},
  \bibinfo{journal}{JHEP} \textbf{\bibinfo{volume}{04}}, \bibinfo{pages}{251}
  (\bibinfo{year}{2021}), \eprint{2012.03713}.

\bibitem[{\citenamefont{Emken et~al.}(2021)\citenamefont{Emken, Frerick, Heeba,
  and Kahlhoefer}}]{Emken:2021vmf}
\bibinfo{author}{\bibfnamefont{T.}~\bibnamefont{Emken}},
  \bibinfo{author}{\bibfnamefont{J.}~\bibnamefont{Frerick}},
  \bibinfo{author}{\bibfnamefont{S.}~\bibnamefont{Heeba}}, \bibnamefont{and}
  \bibinfo{author}{\bibfnamefont{F.}~\bibnamefont{Kahlhoefer}}
  (\bibinfo{year}{2021}), \eprint{2112.06930}.

\bibitem[{\citenamefont{Dror et~al.}(2021)\citenamefont{Dror, Elor, McGehee,
  and Yu}}]{Dror:2020czw}
\bibinfo{author}{\bibfnamefont{J.~A.} \bibnamefont{Dror}},
  \bibinfo{author}{\bibfnamefont{G.}~\bibnamefont{Elor}},
  \bibinfo{author}{\bibfnamefont{R.}~\bibnamefont{McGehee}}, \bibnamefont{and}
  \bibinfo{author}{\bibfnamefont{T.-T.} \bibnamefont{Yu}},
  \bibinfo{journal}{Phys. Rev. D} \textbf{\bibinfo{volume}{103}},
  \bibinfo{pages}{035001} (\bibinfo{year}{2021}), \eprint{2011.01940}.

\bibitem[{\citenamefont{An and Yang}(2021)}]{An:2020tcg}
\bibinfo{author}{\bibfnamefont{H.}~\bibnamefont{An}} \bibnamefont{and}
  \bibinfo{author}{\bibfnamefont{D.}~\bibnamefont{Yang}},
  \bibinfo{journal}{Phys. Lett. B} \textbf{\bibinfo{volume}{818}},
  \bibinfo{pages}{136408} (\bibinfo{year}{2021}), \eprint{2006.15672}.

\bibitem[{\citenamefont{Song et~al.}(2021)\citenamefont{Song, Nagorny, and
  Vincent}}]{Song:2021yar}
\bibinfo{author}{\bibfnamefont{N.}~\bibnamefont{Song}},
  \bibinfo{author}{\bibfnamefont{S.}~\bibnamefont{Nagorny}}, \bibnamefont{and}
  \bibinfo{author}{\bibfnamefont{A.~C.} \bibnamefont{Vincent}},
  \bibinfo{journal}{Phys. Rev. D} \textbf{\bibinfo{volume}{104}},
  \bibinfo{pages}{103032} (\bibinfo{year}{2021}), \eprint{2104.09517}.

\bibitem[{\citenamefont{Bell et~al.}(2021{\natexlab{b}})\citenamefont{Bell,
  Dent, Dutta, Ghosh, Kumar, Newstead, and Shoemaker}}]{Bell:2021xff}
\bibinfo{author}{\bibfnamefont{N.~F.} \bibnamefont{Bell}},
  \bibinfo{author}{\bibfnamefont{J.~B.} \bibnamefont{Dent}},
  \bibinfo{author}{\bibfnamefont{B.}~\bibnamefont{Dutta}},
  \bibinfo{author}{\bibfnamefont{S.}~\bibnamefont{Ghosh}},
  \bibinfo{author}{\bibfnamefont{J.}~\bibnamefont{Kumar}},
  \bibinfo{author}{\bibfnamefont{J.~L.} \bibnamefont{Newstead}},
  \bibnamefont{and} \bibinfo{author}{\bibfnamefont{I.~M.}
  \bibnamefont{Shoemaker}}, \bibinfo{journal}{Phys. Rev. D}
  \textbf{\bibinfo{volume}{104}}, \bibinfo{pages}{076020}
  (\bibinfo{year}{2021}{\natexlab{b}}), \eprint{2108.00583}.

\bibitem[{\citenamefont{Holdom}(1986)}]{Holdom:1985ag}
\bibinfo{author}{\bibfnamefont{B.}~\bibnamefont{Holdom}},
  \bibinfo{journal}{Phys. Lett. B} \textbf{\bibinfo{volume}{166}},
  \bibinfo{pages}{196} (\bibinfo{year}{1986}).

\bibitem[{\citenamefont{Drukier et~al.}(1986)\citenamefont{Drukier, Freese, and
  Spergel}}]{Drukier:1986tm}
\bibinfo{author}{\bibfnamefont{A.~K.} \bibnamefont{Drukier}},
  \bibinfo{author}{\bibfnamefont{K.}~\bibnamefont{Freese}}, \bibnamefont{and}
  \bibinfo{author}{\bibfnamefont{D.~N.} \bibnamefont{Spergel}},
  \bibinfo{journal}{Phys. Rev. D} \textbf{\bibinfo{volume}{33}},
  \bibinfo{pages}{3495} (\bibinfo{year}{1986}).

\bibitem[{\citenamefont{Radick et~al.}(2021)\citenamefont{Radick, Taki, and
  Yu}}]{Radick:2020qip}
\bibinfo{author}{\bibfnamefont{A.}~\bibnamefont{Radick}},
  \bibinfo{author}{\bibfnamefont{A.-M.} \bibnamefont{Taki}}, \bibnamefont{and}
  \bibinfo{author}{\bibfnamefont{T.-T.} \bibnamefont{Yu}},
  \bibinfo{journal}{JCAP} \textbf{\bibinfo{volume}{02}}, \bibinfo{pages}{004}
  (\bibinfo{year}{2021}), \eprint{2011.02493}.

\bibitem[{\citenamefont{Maity et~al.}(2021)\citenamefont{Maity, Ray, and
  Sarkar}}]{Maity:2020wic}
\bibinfo{author}{\bibfnamefont{T.~N.} \bibnamefont{Maity}},
  \bibinfo{author}{\bibfnamefont{T.~S.} \bibnamefont{Ray}}, \bibnamefont{and}
  \bibinfo{author}{\bibfnamefont{S.}~\bibnamefont{Sarkar}},
  \bibinfo{journal}{Eur. Phys. J. C} \textbf{\bibinfo{volume}{81}},
  \bibinfo{pages}{1005} (\bibinfo{year}{2021}), \eprint{2011.12896}.

\bibitem[{\citenamefont{Bovy and Tremaine}(2012)}]{Bovy:2012tw}
\bibinfo{author}{\bibfnamefont{J.}~\bibnamefont{Bovy}} \bibnamefont{and}
  \bibinfo{author}{\bibfnamefont{S.}~\bibnamefont{Tremaine}},
  \bibinfo{journal}{Astrophys. J.} \textbf{\bibinfo{volume}{756}},
  \bibinfo{pages}{89} (\bibinfo{year}{2012}), \eprint{1205.4033}.

\bibitem[{\citenamefont{Rodrigues et~al.}(2021)}]{Rodrigues:2020xpt}
\bibinfo{author}{\bibfnamefont{D.}~\bibnamefont{Rodrigues}}
  \bibnamefont{et~al.}, \bibinfo{journal}{Nucl. Instrum. Meth. A}
  \textbf{\bibinfo{volume}{1010}}, \bibinfo{pages}{165511}
  (\bibinfo{year}{2021}), \eprint{2004.11499}.

\bibitem[{\citenamefont{Xia et~al.}(2022)\citenamefont{Xia, Xu, and
  Zhou}}]{Xia:2021vbz}
\bibinfo{author}{\bibfnamefont{C.}~\bibnamefont{Xia}},
  \bibinfo{author}{\bibfnamefont{Y.-H.} \bibnamefont{Xu}}, \bibnamefont{and}
  \bibinfo{author}{\bibfnamefont{Y.-F.} \bibnamefont{Zhou}},
  \bibinfo{journal}{JCAP} \textbf{\bibinfo{volume}{02}}, \bibinfo{pages}{028}
  (\bibinfo{year}{2022}), \eprint{2111.05559}.

\end{thebibliography}
\end{document}